\documentclass[12pt]{article}

\DeclareSymbolFont{bletters}{OML}{cmm}{bx}{it}
\DeclareMathSymbol{\bdl}{\mathord}{bletters}{"0E}
\DeclareMathSymbol{\bsi}{\mathord}{bletters}{"1B}
\DeclareMathSymbol{\bphi}{\mathord}{bletters}{"1E}
\DeclareMathSymbol{\bbe}{\mathord}{bletters}{"0C}
\DeclareMathSymbol{\bchi}{\mathord}{bletters}{"1F}
\DeclareMathSymbol{\bvphi}{\mathord}{bletters}{'047}
\DeclareMathSymbol{\bPsi}{\mathord}{bletters}{"09}
\DeclareMathSymbol{\bta}{\mathord}{bletters}{'021}

\begin{document}

\title {
$${}$$
{\bf A modified screw dislocation }\\
{\bf with non-singular core of finite radius }\\
{\bf from Einstein-like gauge equation }\\
{\bf (non-linear approach)}
$${}$$}

\author{C. MALYSHEV\\
$${}$$
{\it V. A. Steklov Institute of Mathematics,}\\
{\it St.-Petersburg Department,}\\
{\it Fontanka 27, St.Petersburg, 191023, RUSSIA}\\
E-mail: malyshev@pdmi.ras.ru}

\maketitle

\def \al{\alpha}
\def \be{\beta}
\def \ga{\gamma}
\def \dl{\delta}
\def \ep{\varepsilon}
\def \ze{\zeta}
\def \nb{\nabla}
\def \th{\theta}
\def \ka{\varkappa}
\def \la{\lambda}
\def \si{\sigma}
\def \ph{\varphi}
\def \om{\omega}
\def \Ga{\Gamma}
\def \Dl{\Delta}
\def \La{\Lambda}
\def \Si{\Sigma}
\def \Ph{\Phi}
\def \Om{\Omega}
\def \cA{\cal A}
\def \cB{\cal B}
\def \cC{\cal C}
\def \cD{\cal D}
\def \cE{\cal E}
\def \cN{\cal N}
\def \cI{\cal I}
\def \cR{\cal R}
\def \cF{\cal F}
\def \cY{\cal Y}
\def \cK{\cal K}
\def \cJ{\cal J}
\def \cZ{\cal Z}
\def \BC{I\!\!\!\! C}
\def \BD{I\!\!\!\! D}
\def \BZ{Z\!\!\!Z}
\def \BR{I\!\! R}
\def \IM{\Im}
\def \RE{\Re}
\def \1{^{-1}}
\def \cd{\partial}
\def \cD{{\cal D}}
\def \cA{{\cal A}}
\def \cM{{\cal M}}
\def \cL{{\cal L}}
\def \at{{\rm arctan}\,}
\def \ch{{\rm ch}\,}
\def \sh{{\rm sh}\,}
\def \th{{\rm th}\,}
\def \ld{\ldots}
\def \CU{{\cal U}}
\def \BQ{I\!\!Q\!\!I}
\def \bQ{{\bf Q}}
\def \vt{\vartheta}
\def \w{\widetilde}
\def \h{\widehat}
\def \d{{\rm d}\,}
\def \t{\times}
\def \l{\langle}
\def \r{\rangle}
\def \Tr{{\rm Tr}\,}
\def \Inc{{\rm Inc}\,}
\def \tr{{\rm tr}\,}
\def \diag{{\rm diag}\,}
\def \Det{{\rm Det}\,}
\def \z{\zeta}
\def \iso{{\it iso}(3)}
\def \ISO{{\it ISO}(3)}
\def \T{{\it T}(3)}
\def \S{{\it SO}(3)}
\def \ppu{\times\!\!\!\!\!\!\supset}
\def \pps{+\!\!\!\!\!\!\supset}

\begin{abstract}
\noindent
A continual model of non-singular screw dislocation lying along a 
straight infinitely long circular cylinder is investigated in the 
framework of translational gauge approach with the Hilbert--Einstein 
gauge Lagrangian. The stress--strain constitutive law implies the 
elastic energy of isotropic continuum which includes the terms of 
second and third orders in the strain components. The Einstein-type 
gauge equation with the elastic stress tensor as a driving source is 
investigated perturbatively, and second order contribution
to the stress potential of the modified screw dislocation
is obtained. A stress-free boundary condition is imposed at the 
cylinder's external surface. A cut-off of the classical approach which 
excludes from consideration a tubular vicinity of the defect's axis is 
avoided, and the total stress obtained for the screw dislocation is 
valid in the whole body. An expression for the radius of the 
dislocation's core in terms of the second and third order elastic 
constants is obtained. 
\end{abstract}

\vskip 3.0cm
\rightline{{\bf cond-mat/0312709}}

\newpage

\section{Introduction}

The translational gauge approach based on the Hilbert--Einstein 
gauge Lagrangian has been proposed in \cite{mal} for description
of static dislocations in continual solids. The group of translations 
of three-dimensional space $\T\approx{\BR}^3$ is accepted in 
\cite{mal} as the gauge group. The model \cite{mal} leads,
in linear approximation, to so-called {\it modified} defects
instead of the ordinary dislocation solutions of theoretical 
elasticity. The modified defects demonstrate a non-singular
behaviour, i.e., are characterized by absence of the axial
singularities inherent to the classical screw and edge Volterra 
dislocation solutions. The present paper is devoted to further 
development of the approach \cite{mal}. More specifically, it is
to continue the investigation of the modified screw dislocation
obtained in \cite{mal} and to propose a way of derivation of
second order corrections to its stress field.

The point is that the Einstein-type gauge equation arising
in \cite{mal} to govern the $\T$-gauge fields admits, in 
linear approximation, two short-ranged solutions (so-called, 
{\it modified} or {\it gauge} stress potentials) which coincide 
asymptotically with the stress potentials (i.e., with the Prandtl and 
the Airy stress 
functions) of the ordinary screw and edge dislocations. Accordingly 
to the picture proposed in \cite{mal}, the stress fields calculated
by means of the modified stress potentials just imply additional 
`gauge' contributions to the corresponding stress fields of 
appropriate classical dislocations considered as background
``configurations'' (i.e., as pre-imposed sources of internal 
stresses). In other words, superposition of two stress fields, one is 
due to a chosen classical Volterra dislocation and another is due to 
the corresponding short-ranged gauge stress potential (which is 
localized within a vicinity of the background defect's axis), should 
be considered as the total solution of the gauge model in question.

Therefore, two total solutions obtained in \cite{mal} in the 
super-imposed form are characterized by {\it core region}, where 
singularities of the classical edge and screw dislocations are 
smoothed out. In other words, the gauge approach which is based on 
the Hilbert--Einstein gauge Lagrangian allows to avoid the artificial 
singularities of the classical elasticity. Thus, the gauge approach 
``generates'' a length scale in a continuous description \cite{mal}, 
\cite{laz2}, \cite{ed3}. The length scale characterizes the size of 
the domain where the classical law $1/\rho$ of the dislocation stresses
ceases to be valid and where the axial singularity is ``avoided''. 
Outside such domain the components of the background stresses become 
dominating. Thus, in the framework of \cite{mal} it is possible to 
study the modified defects which allow to reproduce the stresses of 
the classical dislocations ($\si_{\phi z}$ of the screw dislocation, 
and $\si_{\rho \rho}$, $\si_{\rho \phi}$ of the edge dislocation; 
$\si_{\phi \phi}$, $\si_{z z}$ of the modified edge defect \cite{mal} 
behave unconventionally) sufficiently far from their axes while the 
stress components tend to zero within the core regions.

Let us turn to the screw dislocation. Approaches \cite{ks}, 
\cite{pfl}, \cite{seegm}, \cite{wseeg} are known as attempts 
to go beyond the linear elasticity in 
description of the edge and screw dislocations. Non-linear 
approach (second-order elasticity, in fact) can be used to 
find corrections to the rule $1/\rho$. However, it is still 
impossible to approach to the axis of a line defect 
sufficiently close since the conventional theoretical 
elasticity fails. For instance, the fields of second order 
stresses have been found in \cite{ks}, \cite{pfl} (by means of
the stress function method) and in \cite{seegm} (in the 
displacement function approach) which are valid within a hollow 
cylinder with the outer radius, say, $\rho_e$ and the inner 
one $\rho_c$. Free parameters of the models are fixed by 
requirements of stress-free boundaries at $\rho=\rho_e$ and 
$\rho=\rho_c$: $\si_{\rho \rho}|_{\rho=\rho_e}=0$, 
$\si_{\rho \rho}|_{\rho=\rho_c}=0$, where $\si_{\rho \rho}$ is the 
radial stress component (however, the boundary conditions
are written in \cite{pfl} and in \cite{seegm} with respect of
the final and initial states, accordingly). Besides, vanishing of 
$\si_{z z}$ averaged over bulk's cross-section is also used for 
determination of one of the free parameters. Approaches 
mentioned do not consider the region $0\le\rho\le\rho_e$. 

A discussion of relevance of second order effects in theoretical
elasticity for physics of imperfections
in crystals, namely for modelling dislocations, can be found in 
\cite{haif}. Thus, it is clearly interesting to investigate the 
model proposed in \cite{mal} in second order also. This is just 
the problem to be studied in the present paper. However, its 
purely mathematical aspect is of primary interest here.

As to the gauge approaches to defects in continual solids, an 
attempt \cite{osip} is known to follow \cite{ed1} in obtaining 
second order contributions to the stress field of the screw 
dislocation. To this purpose, the quadratic translational 
gauge Lagrangian \cite{ed1} is used in \cite{osip}. However, 
as it is explained in \cite{mal}, the quadratic $\T$-gauge Lagrangian 
advanced in \cite{ed1}, \cite{ed2} is inappropriate since it 
forbids a modified stress potential which correctly reproduces 
the stress field of the edge dislocation. From the point of view of 
the Refs. \cite{laz1}, \cite{laz2} \footnote{A gauge approach close 
to ours is proposed in \cite{laz1}, \cite{laz2} which is based on the 
translational gauge Lagrangian ${\cal L}_T$ written as a combination 
of terms quadratic in the torsion components (i.e., in the dislocation 
density's components). For a special choice of the parameters, 
${\cal L}_T$ is equivalent to the Hilbert-Einstein Lagrangian 
\cite{mal}. After \cite{kat} it is known that extension of the 
Hilbert--Einstein Lagrangian by terms quadratic in torsion (and 
curvature) leads to quadratic in torsion Lagrangians of more 
general form (as well as to the most general eight-parameter three
dimensional Lagrangian \cite{kat}).}, the Lagrangian used in 
\cite{osip}, being considered as a form quadratic in the torsion 
components, is incomplete. Besides, the elastic energy is also taken in 
\cite{osip} in a restricted (in comparison with that of the classical, 
i.e., non-gauge approaches \cite{ks}, \cite{pfl}, \cite{seegm}) form. 
Since the gauge Lagrangian in \cite{osip} is inappropriate to 
capture the edge dislocation, it is also insufficient to consider second 
order corrections to the screw dislocation: the Kr\"oner ansatz for the 
second order stresses of the screw dislocation is just of the same form 
as that used for an edge dislocation. Thus, the experience of 
\cite{osip} looks unsatisfactory.

The present paper is to demonstrate that second order 
consideration can be carried out along the line of the classical 
investigation \cite{pfl} for the gauge model proposed in \cite{mal} 
also. Namely, we shall consider the second order solution found in 
\cite{pfl} for the straight screw dislocation lying along cylindric 
body as a background source of internal stresses. In this case, 
solution of the Einstein-type gauge equation gives a short-ranged 
``correction'' to the classical background. The short-ranged gauge 
solution depends on several free parameters. We shall adjust these 
parameters in a way which differs from that in \cite{pfl}, 
\cite{seegm}: for instance, the vanishing boundary condition will be 
imposed only for the outer surface of cylindric body containing the 
dislocation. Instead, we shall require vanishing of certain 
coefficients in the short distance expansion of the second order 
stress potential.

Specifically, it will be demonstrated that one of the
`matching conditions' for the free parameters results in an 
expression which relates the radius of the domain of localization 
of the defect's density profile to some second and third order elastic 
moduli. The main statement of the paper thus reads: the second order 
solution obtained demonstrates that singularities in $\si_{\rho \rho}$, 
$\si_{\phi \phi}$ do not appear, and the 
stress components tend either to zero or to constant values
(accordingly to the choice of the free parameters) at $\rho\to 0$. 
However, $\si_{z z}$ is still weakly (i.e., logarithmically) divergent, 
though it is integrable over the cylinder's cross-section.
Thus the value for $\si_{z z}$ averaged over the cross-section 
surface is finite. The week divergency of $\si_{z z}$ is due
to a simplifying assumption about the defect's density profile.

In the present paper we are not to discuss in specific
details such a complicated field as description of the core 
structures of the crystal dislocations. Instead,
only a list (inevitably incomplete) of further references
is proposed: for instance, one should be referred to 
\cite{pei}, \cite{nab}, \cite{merw}, \cite{marad}
for the first attempts to incorporate discreteness for consideration
of the core structures. Further references, say, for (non-linear) 
elasticity, for crystallography, for discrete (atomic) and mixed 
approaches, etc., can be found also in \cite{kott}, \cite{hir}, 
\cite{teod}, \cite{gair}. For theory and experiment concerning
the dislocation core structures and for effects of influence of 
the dislocation core structures on various physical properties
of solids one should be referred to \cite{yam}, \cite{cnrs}
(besides, certain refs. omitted below should be found in 
\cite{mal}).

The paper is written in six sections. Section 1 is 
introductory (see also \cite{mal} for motives of our approach and 
for appropriate refs.). Section 2 is to outline the Einstein-type 
gauge equation. Section 3 is devoted to further specifications 
of the gauge equation, and a perturbative scheme is set up. Solution 
to the gauge equation which describes the modified stress potentials 
of second order is obtained in Section 4. The corresponding
components of the stress field and their asymptotics are investigated 
in Section 5. Discussion in Section 6 concludes the paper. Details 
of the calculation are provided in Appendices A and B.
Bold-faced letters are used to denote tensors of second rank
(i.e., loosely speaking, matrices).

\section{The Hilbert--Einstein gauge equation     }

The aim of the present paper is to deduce the stress field
of second order of the modified screw dislocation obtained, 
in linear approximation, in \cite{mal}. It should be reminded 
that before than in \cite{mal}, the modified screw dislocation 
was already obtained in \cite{ed3} for a translational gauge 
model based on the gauge Lagrangian quadratic in the torsion 
components (a `restricted' choice of the gauge invariant 
quadratic form). Besides, the same modified screw dislocation 
was reproduced also in \cite{laz2}, \cite{laz3} for the gauge 
Lagrangian taken as a more general (in comparison with \cite{ed3}) 
quadratic form. For an appropriate choice of the parameters, 
the Lagrangian \cite{laz2}, \cite{laz3} is equivalent to the 
Hilbert--Einstein Lagrangian proposed in \cite{mal}. It should be 
pointed out that the same, i.e., like in \cite{ed3} and \cite{laz2}, 
non-singular screw dislocation first, seemingly, appeared in 
\cite{cem} in the framework of non-local elasticity approach.

The main idea behind all these gauge attempts, \cite{ed3}, \cite{mal}, 
\cite{laz2}, can be summarized as follows. Conventionally, 
the ordinary dislocations are characterized by the stress tensors 
$\bsi$ which are singular on the defect's axes. In the gauge 
approaches mentioned, additional gauge contributions to the 
stress fields appear so that within compact regions (the core 
regions) the classical singularities are smoothed out. At sufficiently 
large distances, the stress components of the modified defects 
demonstrate the behaviour inherent to the classical dislocations. 
Within the cores transition between two asymptotics occures.

We are going to consider the model proposed in \cite{mal}, and
second order elasticity approach is accepted below. In the 
present section, the Einstein-type gauge equation \cite{mal} 
is outlined. Some differential--geometric notations are reminded, 
but for more details about them one should refer to \cite{gair}, 
\cite{lh}, \cite{klein}. It is important that now we are using 
the Eulerian picture instead of the Lagrangian one accepted in 
\cite{mal}. The Lagrangian and the Eulerian pictures are 
indistinguishable in the linear approximation.  

Our picture is based on the Eulerian strain tensor \cite{gair} 
related to deformed (final) state of a dislocated body. Let 
us denote the squared length element between two neighboring 
points before deformation as $dS^2$, and the squared length in a 
final state will be denoted as $ds^2$. We consider the difference 
between $ds^2$ and $dS^2$, and thus we introduce the Eulerian 
strain tensor $e_{a b}$ as follows.

Let us introduce the triples $\{x^i\}$ and $\{x^a\}$ as the
coordinates bases (Cartesian or curvilinear) to be used 
for description of initial and final states, respectively. The 
corresponding squared length elements can be written as
$g_{i j} dx^idx^j$ or $\eta_{a b} dx^a dx^b$
(with $\eta_{a b}\equiv g_{i j} {\cE}_a^{\,\,i}{\cE}_b^{\,\,j}$)
with respect to $\{x^i\}$ or $\{x^a\}$, accordingly.
Let us define the frame components by means of the relation
$\cd_{i}=e^a_{\,\,i}\cd_{a}$ (here and below partial derivatives
$\cd/\cd x^i$ are denoted as $\cd_i$), and co-frame components 
${\cE}_a^{\,\,i}$ -- by means of the one-form 
$dx^i={\cE}_a^{\,\,i} dx^a$. The components ${\cE}_a^{\,\,i}$ with 
their duals $e^a_{\,\,i}$ are orthogonal in the following sense:
$$
e^a_{\,\,i} {\cE}_b^{\,\,i}=\dl^a_b\,,\qquad e^a_{\,\,i} 
{\cE}_a^{\,\,j}=\dl^j_i
$$ 
(throughout the paper repeated indices imply summation). Further, 
let us consider a map from an initial state to the deformed state 
$\{\xi^a\}$ as follows:
$$\xi\,:\,\,x^i\,\longmapsto\,\xi^a(x^i)\,.
$$
Consideration of the difference
$$
ds^2-dS^2=\eta_{a b} d\xi^a d\xi^b-g_{i j} dx^idx^j=
2 e_{a b}d\xi^a d\xi^b, 
\eqno(2.1)
$$
where
$$
2 e_{a b}\equiv \eta_{a b} - g_{a b}, 
\qquad g_{a b}\equiv g_{i j} {\cB}_a^{\,\,i}{\cB}_b^{\,\,j}\,,
\eqno(2.2)
$$
allows to define the Eulerian strain tensor $e_{a b}$.
Here ${\cB}_a^{\,\,i}$ are the coefficients in
$1$-forms $dx^i={\cB}_a^{\,\,i} d\xi^a$. The metric tensor 
$g_{a b}$ (2.2) is called the Cauchy deformation tensor.

In the absence of defects, ${\cB}_a^{\,\,i}$ is expressed as 
follows \cite{gair}:
$$
{\cB}_a^{\,\,i}\,\equiv\,\frac{\cd x^i}{\cd\xi^a}\,=\,
{\cE}_a^{\,\,i}\,-\,{\cE}_b^{\,\,i}\,
\stackrel{(\eta)}{\nabla}_a u^b\,,
\eqno(2.3.1)
$$
where $\xi^i=x^i+u^a {\cE}_a^{\,\,i}$ with respect to the 
base $\{x^i\}$. The covariant derivative
$\stackrel{(\eta)}{\nabla}_a$ in (2.3.1) is defined by the
requirement that the components ${\cE}_a^{\,\,i}$ are
covariantly constant, i.e.,
$$
\stackrel{(\eta)}{\nabla}_a\,{\cE}_b^{\,\,i}\,\equiv\,
\cd_a {\cE}_b^{\,\,i}\,-\,
\left\{\begin{array} {cc} c \\a b\end{array}
                 \right\}_\eta\,{\cE}_c^{\,\,i}\,=\,0\,,
\eqno(2.3.2)
$$
and thus the metric $\eta_{a b}={\cE}_a^{\,\,i}{\cE}_{b\,i}$
is covariantly constant. With the help of (2.3.2), we can express 
the Christoffel symbol of second kind 
$\left\{\begin{array} {cc} c \\a b\end{array} 
\right\}_\eta$ through the metric $\eta_{a b}$ as follows:
$$
\left\{\begin{array} {cc} c \\a b\end{array} 
\right\}_\eta\,=\,
\displaystyle{\frac 12}\,\eta^{c e} \left(
\cd_a \eta_{b e}\,+\,
\cd_b \eta_{a e}\,-\,\cd_e \eta_{a b}\right)\,.
\eqno(2.3.3)
$$

In the presence of translational defects we put new ${\cB}_a^{\,\,i}$
in another, in comparison with (2.3), form:
$$
{\cB}_a^{\,\,i}\,\equiv\,\frac{\cd x^i}{\cd\xi^a}
\,-\,\ph_a^{\,\,i}\,=\,
{\cE}_a^{\,\,i}\,-\,\left({\cE}_b^{\,\,i}\,
\stackrel{(\eta)}{\nabla}_a u^b\,+\,\ph_a^{\,\,i}\right)\,.
\eqno(2.4)
$$
Here $\ph_a^{\,\,i}$ are the translational gauge potentials 
which are transformed under a non-homogeneous $T(3)$-gauge
transformation $x^i\,\longrightarrow\,x^i+\eta^i(x)$ as follows:
$$
\begin{array}{lcr}
\displaystyle{
\frac{\cd x^i}{\cd\xi^a}}&\longrightarrow&\displaystyle{
\frac{\cd x^j}{\cd\xi^a}\,\Bigl(\dl^i_j\,+\,\frac{\cd \eta^i}{\cd x^j}
\Bigr)}\,,\\
[0.5cm] 
\displaystyle{
\ph_a^{\,\,i}}&\longrightarrow&\displaystyle{\!\!\!\!\!\!\!
\ph_a^{\,\,i}\,+\,\frac{\cd x^j}{\cd\xi^a}\,
\frac{\cd \eta^i}{\cd x^j}}\,.
\end{array}
\eqno(2.5)
$$
The replacements (2.5) ensure the gauge invariance of the components 
${\cB}_a^{\,\,i}$. I.e., in the presence of defects, ${\cB}_a^{\,\,i}$ 
behave like $\cd x^i/\cd\xi^a$ (2.3.1) at homogeneous 
($\cd \eta^i/\cd x^j\equiv 0$) transformations.

Eventually, when (2.4) takes place, the strain $e_{a b}$ (2.2) can 
be written as follows:
$$
2 e_{a b}\,=\,\stackrel{(\eta)}{\nabla}_a u_b\,+\,\ph_{a b}
\,+\,\stackrel{(\eta)}{\nabla}_b u_a\,+\,\ph_{b a}
\,-\,
\left(\stackrel{(\eta)}{\nabla}_a u_b\,+\,\ph_{a b}\right)
\left(\stackrel{(\eta)}{\nabla}_b u_a\,+\,\ph_{b a}\right)
\,,
\eqno(2.6)
$$
where the components $\ph_{a b}$ of the gauge potential $\bvphi$ are 
defined by means of the representation 
$\ph_b^{\,\,i}\,=\,{\cE}_a^{\,\,i}\,\ph_b^{\,\,a}$ provided $\eta_{a b}$ 
is used to rise and lower the indices. When the gauge potential $\bvphi$ 
is zero, Eq. (2.6) is reduced to conventinal expression for the strain 
${\bf e}$ in the Eulerian picture \cite{gair}.

Further, in order to consider the gauge equation, we need the 
Riemann--Christoffel curvature tensor 
${{\rm R}_{a b c}}^d\,\,$:
$$
{{\rm R}_{a b c}}^d
\,=\,
\cd_a \left\{\begin{array} {cc} d \\b c\end{array}
                                        \right\}_g\,-\,
\cd_b \left\{\begin{array} {cc} d \\a c\end{array}
                                        \right\}_g\,+\,
\left\{\begin{array} {cc} d \\a e\end{array}
                                         \right\}_g
\left\{\begin{array} {cc} e \\b c\end{array}
                                         \right\}_g\,-\,
\left\{\begin{array} {cc} d \\b e\end{array}
                                         \right\}_g
\left\{\begin{array} {cc} e \\a c\end{array}
                                          \right\}_g\,,
\eqno(2.7)
$$
where
$$
\begin{array}{l}
\displaystyle{
\left\{\begin{array} {cc} d \\b c\end{array}
                                          \right\}_g
\equiv \left\{b c, e\right\}_g\,g^{e d}}\,,\\ [0.6cm]
\{b c, e\}_g\,\equiv\,\displaystyle{\frac 12} 
\left(\cd_b g_{c e}\,+\,
\cd_c g_{b e}\,-\,\cd_e g_{b c}\right)\,.
\end{array}
\eqno(2.8)
$$
In (2.7) and (2.8), the Christoffel symbols of first and 
second kind, $\{b c, e\}_g$ and  
$\displaystyle{\left\{\begin{array} {cc} d \\b c\end{array}
\right\}_g}$, accordingly, are calculated with
respect to the metric $g_{a b}$ (2.2) (the subscript `g').
The metric $g_{a b}$ is covariantly constant with respect
to the metric connection $\stackrel{(g)}{\nabla}$ which is 
expressed through the Christoffel symbols (2.8), i.e., equation
$\stackrel{(g)}{\nabla}_a g_{b c}=0$ is fulfilled.

Since the metric $\eta_{a b}$ is also covariantly constant 
in terms of the corresponding metric connection 
$\stackrel{(\eta)}{\nabla}$,
$$
\stackrel{(\eta)}{\nabla}_a \eta_{b c}= 0\,,
\eqno(2.9)
$$
we obtain from (2.9) with the help of (2.2):
$$
\cd_a g_{b c}=
\left\{\begin{array} {cc} e \\a b\end{array}
                 \right\}_\eta g_{e c}
+
\left\{\begin{array} {cc} e \\a c\end{array}
                 \right\}_\eta g_{b e}
-2 \stackrel{(\eta)}{\nabla}_a e_{b c}\,,
\eqno(2.10)
$$
where the Christoffel symbols of second kind are defined by means
of (2.3.3). In its turn, Eq.(2.10) (plus two other
equations due to cyclic permutations of the indices) 
gives us:
$$
\begin{array}{l}
\displaystyle{
\left\{\begin{array} {cc} c \\a b\end{array}
                 \right\}_g =
\left\{\begin{array} {cc} c \\a b\end{array}
                 \right\}_\eta 
-2 {e_{a b}}^c}\,,\\[0.8cm]
2 {e_{a b}}^c\equiv g^{c e}\Bigl(
\stackrel{(\eta)}{\nabla}_a e_{b e}
+\stackrel{(\eta)}{\nabla}_b e_{a e}
-\stackrel{(\eta)}{\nabla}_e e_{a b}\Bigr)\,.
\end{array}
\eqno(2.11)
$$

We substitute (2.11) into (2.7) and obtain another 
representation for the curvature tensor:
$$
{{\rm R}_{a b c}}^d = 
{\stackrel{(\eta)}{{\rm R}}_{a b c}}{}^{d}
-2 \Biggl(\stackrel{(\eta)}{\nabla}_a {e_{b c}}^d
-2 {e_{a e}}^d {e_{b c}}^e\,-\,
\Bigl(a\longleftrightarrow b \Bigr) \Biggr)\,,
\eqno(2.12)
$$
where ${\stackrel{(\eta)}{{\rm R}}_{a b c}}{}^{d}$ is the Riemann 
curvature calculated analogously to (2.7) but for the metric 
$\eta_{a b}$. In our Eulerian approach, we assume that the 
geometry of the deformed state is flat, and so we put 
${\stackrel{(\eta)}{{\rm R}}_{a b c}}{}^{d}$ equal to zero. Now we 
are ready to write the Hilbert--Einstein gauge equation. Let us 
define the Einstein tensor $G^{e f}$ as follows:
$$
G^{e f}\,=\,\frac 14\,{\cE}^{e a b} {\cE}^{f c d} {\rm R}_{a b c d}\,,
\eqno(2.13)
$$
or, with the help of (2.12),
$$
G^{e f}\,=\,-{\cE}^{e a b} {\cE}^{f c d} 
\stackrel{(\eta)}{\nabla}_a\,\stackrel{(\eta)}{\nabla}_c
\,e_{b d}\,-\,2 {\cE}^{e a b} {\cE}^{f c d} e_{a d e^{\prime}}
\,{e_{b c}}^{e^{\prime}}\,.
\eqno(2.14)
$$
In (2.13) and (2.14), ${\cE}^{a b c}$ is the totally antisymmetric
Levi--Civita tensor \cite{gair} defined by means of the metric 
$\eta_{a b}$. Therefore, the gauge equation proposed in \cite{mal} 
(Section 6) takes the form:
$$
G^{e f}\,=\,(2 s)^{-1} \Bigl(\si^{e f}\,-
\,(\si_{{\rm bg}})^{e f}\Bigr)\,,
\eqno(2.15)
$$
or,
$$
-{\cE}^{e a b} {\cE}^{f c d} 
\stackrel{(\eta)}{\nabla}_a\,\stackrel{(\eta)}{\nabla}_c
\,e_{b d}\,=\,
(2 s)^{-1} \Bigl(\si^{e f}\,-\,(\si_{{\rm bg}})^{e f}\Bigr)
\,+\,2 {\cE}^{e a b} {\cE}^{f c d} e_{a d e^{\prime}}
\,{e_{b c}}^{e^{\prime}}\,.
\eqno(2.16)
$$

Variational derivation of (2.15) can be discussed along the
line of \cite{mal} where the gauge approach was developed in the 
Lagrangian coordinates. Right-hand side of (2.15) is given by the 
difference ${\bsi}-{\bsi}_{{\rm bg}}$, where ${\bsi}_{{\rm bg}}$ 
implies the stress tensor of a {\it background} defect. The 
difference ${\bsi}-{\bsi}_{{\rm bg}}$, i.e., just the deviation of 
the total stress from ${\bsi}_{{\rm bg}}$, plays the role of the 
source of geometric configurations described by the Einstein 
tensor ${\bf G}$. The parameter $s$ (a {\it coupling constant}, 
accordingly to gauge terminology) characterizes an energy scale 
intrinsic to the gauge field ${\bvphi}$, and it appears 
as a factor at the Hilbert--Einstein Lagrangian density \cite{mal}.

In the present paper, ${\bsi}_{{\rm bg}}$ is assumed to be
given by the stress field of a single straight screw 
dislocation lying along an infinitely long cylindric body.
Practically, the solution provided by \cite{pfl} will be adopted,
which is valid within a hollow cylinder restricted by two 
surfaces: internal ($\rho=\rho_c$) and external 
($\rho=\rho_e$). Further comments about (2.15), (2.16), and 
about their specification for the present problem can be found 
below. Besides, both ${\bsi}$ and ${\bsi}_{{\rm bg}}$ are 
assumed to respect the equilibrium equations:
$$
\stackrel{(\eta)}{\nabla}_a \si^{a b}=0,\qquad
\stackrel{(\eta)}{\nabla}_a (\si_{{\rm bg}})^{a b}=0\,.
\eqno(2.17)
$$

More detailed consideration of the gauge geometry behind the model 
to be studied in the present paper should be done elsewhere. However, 
Refs. \cite{sard} should be mentioned in addition to those listed in 
\cite{mal} since they contain reviewing notes and useful refs. 
concerning translational gauge geometry. It is also interesting to 
note Ref. \cite{riv} where a topological picture is build up which 
includes dislocations (torsion) and exta-matter (non-metricity).

\section{Specification of the gauge equation}
\subsection{The stress function method}

We shall investigate Eq. (2.16) using the method of stress functions 
proposed in \cite {ks} for solving the internal stress problems in 
incompatible elasticity. Specifically, we shall follow Ref. \cite{pfl}, 
where the approach \cite{ks} was further developed in the successive 
approximation form to determine the second order stress fields of the 
screw and edge straight dislocations lying along cylindrical tubes of 
circular cross sections. An exposition of \cite{pfl} can be found in 
Ref. \cite{gair} devoted to a review of dislocation problems
in non-linear elasticity. Certain details (concerning, for instance, 
the tensor formalism in curvilinear coordinates) omitted in what 
follows can be restored with the help of \cite{pfl} and 
\cite{gair}. In what follows, bold-faced letters denote tensors, 
and all the indices can easily be restored.

Accordingly to Eq. (2.6), let us represent the strain and the stress
tensors as the perturbative expressions:
$$
{\bf e}\,=\,\stackrel{(1)}{{\bf e}}\,+\,\stackrel{(2)}{{\bf e}}\,,
\qquad\qquad {\bsi}\,=\,\stackrel{(1)}{{\bsi}}\,+\,
\stackrel{(2)}{{\bsi}}\,,
\eqno(3.1)
$$
where $\stackrel{(2)}{{\bf e}}$, $\stackrel{(2)}{{\bsi}}$ are
assumed to be of second order smallness in comparison to
$\stackrel{(1)}{{\bf e}}$, $\stackrel{(1)}{{\bsi}}$. Expressions
(3.1) can be understood as the first two terms of formal perturbative
series in powers of a small parameter. Substituting (3.1) into (2.16) 
and (2.17), we obtain the following two sets of the governing equations:

\vskip 0.4cm
\noindent
(first order)
$$
\begin{array}{l}
\displaystyle{
\nabla_a \stackrel{(1)}{\si}{}^{a b}=0}\,,\\[0.4cm]
\Bigl({\Inc} \stackrel{(1)}{{\bf e}}\Bigr)^{a b}\,=\,
(2 s)^{-1}\Bigl(\dl \stackrel{(1)}{{\bsi}}\Bigr)^{a b};
\end{array}
\eqno(3.2)
$$

\vskip 0.4cm
\noindent
(second order)
$$
\begin{array}{l}
\displaystyle{
\nabla_a \stackrel{(2)}{\si}{}^{a b}=0}\,,\\[0.4cm]
\Bigl({\Inc} \stackrel{(2)}{{\bf e}}\Bigr)^{a b}\,=\,
(2 s)^{-1}\Bigl(\dl \stackrel{(2)}{{\bsi}}\Bigr)^{a b}\,
+\,\stackrel{(1)}{Q}{}^{a b}\,.
\end{array}
\eqno(3.3)
$$
The following notations are used in (3.2) and (3.3) 
($i=1, 2$):
$$
\Bigl({\Inc} \stackrel{(i)}{{\bf e}}\Bigr)^{a b}
\equiv -{\cal E}^{a c d} {\cal E}^{b f e}
\nabla_c\,\nabla_f\stackrel{(i)}{e}_{d e}\,,
\eqno(3.4.1)
$$
$$
\Bigl(\dl \stackrel{(i)}{{\bsi}}\Bigr)^{a b}
\equiv\,\stackrel{(i)}{\si}{}^{a b} -
\Bigl(\stackrel{(i)}{{\bsi}}_{{\rm bg}}\Bigr)^{a b}\,,
\eqno(3.4.2)
$$
$$
\stackrel{(1)}{Q}{}^{a b}\equiv 
2\,{\cal E}^{a c d} {\cal E}^{b f e}
\stackrel{(1)}{e}_{c e l} \stackrel{(1)}{e}_{d  f}{}^l\,,
\eqno(3.4.3)
$$
where `${\Inc}$' denotes, so-called, {\it incompatibility} 
{\it operator} (acting on a tensor argument), ${\cal E}^{a b c}$ 
is the Levi--Civita tensor, and $\nabla_a$ implies the 
covariant derivative $\stackrel{(\eta)}{\nabla}_a$.
Indices in (3.2)--(3.4) are raised and lowered by means of
the metric $\eta_{a b}$. A direct comparison, for instance, with Eqs. 
(622) and (623) provided by \cite{gair} demonstrates that the
second equations in each pair (3.2) and (3.3) imply transparent 
modification of the corresponding equations of the conventional 
approach: the only difference is given by the terms 
$(2 s)^{-1} \dl \stackrel{(i)}{{\bsi}}$, which are responsible 
for the short-ranged behaviour of the resulting stress functions. 
Besides, the contributions due to the contortion (and thus due 
to the dislocation density) are absent (see explanations in 
\cite{mal}).

The elastic energy potential is chosen in the Eulerian 
representation as follows \cite{pfl}:
$$
W({\bf e})\,=\,j I^2_1 ({\bf e})\,+\,
k I_2 ({\bf e})\,+\,l^{\prime} I^3_1 ({\bf e})
\,+\,m^\prime I_1 ({\bf e})\,I_2 ({\bf e})\,+
\,n^{\prime} I_3 ({\bf e})\,,
\eqno(3.5)
$$
where $j=\mu +\la/2$, $k=-2\mu$ ($\la$ and $\mu$
are the Lam\'e constants), while $l^{\prime}, m^{\prime}, 
n^{\prime}$ are the elastic moduli of third order.
For the given choice of the potential $W({\bf e})$,
the constitutive law which relates $\stackrel{(i)}{\bf e}$
to $\stackrel{(i)}{\bsi}$ takes the following form \cite{pfl}:
$$
\begin{array}{l}
\displaystyle{
\stackrel{(1)}{\bf e}\,=\,
C_1 I_1\bigl(\stackrel{(1)}{\bsi}\bigr) {\bta}
               \,+\,C_4 \stackrel{(1)}{\bsi}}\,,
\\[0.4cm]
\stackrel{(2)}{\bf e}\,=\,
C_1 I_1\bigl(\stackrel{(2)}{\bsi}\bigr) {\bta}
               \,+\,C_4 \stackrel{(2)}{\bsi}
\,+\,\stackrel{(1)}{\bPsi}  \,,
\end{array}
\eqno(3.6)
$$
where
$$
\stackrel{(1)}{\bPsi}\,\equiv 
\Bigl( C_2 I^2_1\bigl(\stackrel{(1)}{\bsi}\bigr)\,+\,
C_3 I_2\bigl(\stackrel{(1)}{\bsi}\bigr)\Bigr){\bta}
\,+\,C_5 I_1\bigl(\stackrel{(1)}{\bsi}\bigr)
\stackrel{(1)}{\bsi}\,+
\,C_7 I_3\bigl(\stackrel{(1)}{\bsi}\bigr) 
\Bigl(\stackrel{(1)}{\bsi}\Bigr)^{-1}\,,
\eqno(3.7)
$$
and the numerical coefficients are \cite{pfl}:
$$
\begin{array}{l}
\displaystyle{
C_1\equiv-\frac\nu E\,, \quad C_4\equiv\frac{1+\nu}E\,,
}\\ [0.5cm]
\displaystyle{
C_2\equiv\frac1{E^2}\bigl(3\nu(1-\nu)-1\bigr)+3L+M\,,
}\\[0.5cm]
\displaystyle{
C_3\equiv\frac3{E^2}\bigl(1-\nu^2\bigr)+M\,,
}\\ [0.5cm]
\displaystyle{
C_5\equiv\frac1{E^2}\bigl(3\nu-2\bigr)\bigl(1+\nu\bigr)
-M\,,
} \\ [0.5cm]
\displaystyle{
C_7\equiv-\frac1{E^2}\bigl(1+\nu\bigr)^2+N}\,.
\end{array}
\eqno(3.8)
$$

In (3.8) we are using the Poisson ratio $\nu$ and the elastic 
modulus $E$: 
$$
\nu=\frac{\,\la\,}{2(\la+\mu)}\,,\qquad
E=\frac{\,\mu(3\la+2\mu)\,}{\la+\mu};
$$
besides, the relationship between the elastic parameters of
third order $L$, $M$, $N$ in (3.8) and $l^\prime$, $m^\prime$, 
$n^\prime$ in (3.5) is given as follows:
$$
\begin{array}{l}
\displaystyle{
n^\prime\,+\,8\mu^3 N\,=\,6 \mu\,, 
}\\[0.5cm]
\displaystyle{
3 m^\prime\,+\,n^\prime\,+\,12 \mu^2 {\cal K}
\bigl(3 M+N \bigr)\,=\,18 {\cal K} \frac{3\nu-2}{1+\nu}\,, 
}\\[0.5cm]
\displaystyle{
27 l^\prime\,+\,9 m^\prime\,+\,n^\prime\,+\,
27{\cal K}^3\bigl(27 L+9 M+N\bigr)\,=\, 36 {\cal K}}\,,
\end{array}
\eqno(3.9)
$$
where
$$
{\cal K}\,\equiv\,\frac{E}{3(1-2\nu)}\,=\,\la\,+\,
\frac23\,\mu\,.
$$
Besides, $\bta$ implies the metric $\eta_{a b}$ in (3.5)--(3.7),
and the functions $I_m$ $(m=1, 2, 3)$ of a tensor argument, say, ${\bf t}$
are defined as follows:
$$
\begin{array}{l}
I_1({\bf t})\,\equiv\,\tr({\bf t})\,,\\[0.4cm]
I_2({\bf t})\,\equiv\,\displaystyle{\frac12}\Bigl(
I_1^2({\bf t})\,-\,I_1({\bf t}^2)\Bigr)\,,\\[0.4cm]
I_3({\bf t})\,\equiv\,\Det ({\bf t})\,.
\end{array}
\eqno(3.10)
$$
The Cayley--Hamilton theorem must be used to express the inverse
$\Bigl(\stackrel{(1)}{\bsi}\Bigr)^{-1}$ in (3.7). More details 
about derivation of Eqs. (3.6)--(3.9) can be found in \cite{pfl}, 
\cite{gair}.

Now we are in position to use the first and the second
equations (3.6) in (3.2) and (3.3), respectively. We obtain:

\vskip 0.4cm
\noindent
(first order)
$$
\begin{array}{l}
\displaystyle{
\nabla_a \stackrel{(1)}{\si}{}^{a b}=0}\,,\\[0.4cm]
\Delta \stackrel{(1)}{\si}{}^{a b}\,+\,
(1-a) \bigl(\nabla_a\nabla_b\,-\,\eta_{a b}\Delta\bigr)
I_1(1)\,=\,\kappa^2 \Bigl(\dl 
           \stackrel{(1)}{{\bsi}}\Bigr)^{a b}\,;
\end{array}
\eqno(3.11)
$$

\vskip 0.4cm
\noindent
(second order)
$$
\begin{array}{lll}
\displaystyle{
\nabla_a \stackrel{(2)}{\si}{}^{a b}=0}\,,& &\\[0.4cm]
\displaystyle{
\Delta \stackrel{(2)}{\si}{}^{a b}\,+\,
(1-a)} \bigl(\nabla_a\nabla_b\,&-&\,\eta_{a b}\Delta\bigr)
I_1(2) \\[0.4cm]&=&\,
\kappa^2 \Bigl(\dl \stackrel{(2)}{{\bsi}}\Bigr)^{a b}\,
+\,2\mu\stackrel{(1)}{Q}{}^{(a b)}
\,-\,2\mu\Bigl({\Inc}\stackrel{(1)}{\bPsi}\Bigr)^{a b}\,,
\end{array}
\eqno(3.12)
$$
where $\stackrel{(1)}{\bPsi}$ is given by (3.7),
$I_1 (i)\equiv I_1 \bigl(\stackrel{(i)}{{\bsi}}\bigr)$, 
$\Delta\equiv \nabla_a\nabla^a$, and
$\stackrel{(1)}{Q}{}^{(a b)}$ is expressed by means of
(3.4.3) and (3.6). Besides, we make use of the parameters
$\kappa^2 \equiv \mu/s$ and $a$ \cite{ed1}, \cite{ed2}:
$$
a\equiv\frac{\,\la\,}{3\la+2\mu}
\,=\,\frac{\,1\,}{1+{\nu}^{-1}}\,,\qquad
1-a=\frac{\,2(\la+\mu)\,}{3\la+2\mu}
\,=\,\frac{\,1\,}{1+\nu}\,.
$$
The curly brackets around the indices imply symmetrization.

We use the stress function ansatz to fulfil the equilibrium 
equations for stresses given by (3.11) and (3.12) as follows:
$$
\stackrel{(i)}{{\bsi}}\,=\,{\Inc} \stackrel{(i)}{{\bchi}}.
\eqno(3.13)                      
$$
Substituting (3.13) into the second equations in (3.11), 
(3.12), we obtain a couple of equations to determine 
the stress potentials $\stackrel{(i)}{{\bchi}}$:
$$
\begin{array}{c}
\displaystyle{
\Dl \Dl \stackrel{(i)}{{\chi}}_{a b}\,+\,
a\Bigl(\nb_a\nb_b\,-\,\eta_{a b}\Dl\Bigr)
 \Dl I_1\bigl(\stackrel{(i)}{{\bchi}}\bigr)\,+\,
\Bigl(\bigl(1-a\bigr)\nb_a\nb_b\,+\,a\,\eta_{a b}
\Dl \Bigr)\nb^c\nb^d \stackrel{(i)}{{\chi}}_{c d}}\,-\\[0.5cm]
-\,\Dl \Bigl(\nb_a\nb_c \stackrel{(i)}{{\chi}}{}^c_{\,\,\,b}
\,+\,\nb_b\nb_c \stackrel{(i)}{{\chi}}{}^c_{\,\,\,a}\Bigr)
\,=\,\kappa^2 \Bigl(\dl \stackrel{(i)}{{\bsi}}\Bigr)_{a b}\,
+\,2\mu\stackrel{(i)}{S}{}_{(a b)}\,,\qquad i=1, 2\,,
\end{array}
\eqno(3.14)
$$
where
$$
\stackrel{(1)}{S}{}_{(a b)}\,\equiv\,0\,,\qquad
\stackrel{(2)}{S}{}_{(a b)}\,\equiv\,
\stackrel{(1)}{Q}{}_{(a b)}\,-\,
\Bigl({\Inc}\stackrel{(1)}{\bPsi}\Bigr)_{(a b)}\,,
$$
and $\dl \stackrel{(i)}{{\bsi}}$ is written by means of (3.4.3)
and (3.13).

Using a linear transformation to another stress potential
$\stackrel{(i)}{{\bchi}}{}^\prime$,
$$
\stackrel{(i)}{{\bchi}}\,=\,
\stackrel{(i)}{{\bchi}}{}^\prime\,+\,
\frac\nu{1-\nu}\,\,\eta_{a b}\,I_1\bigl(
\stackrel{(i)}{{\bchi}}{}^\prime\bigr)\,,
$$
we can reduce (3.14) at $i=2$ to a more simple form:
$$
\Dl\Dl \stackrel{(2)}{{\bchi}}{}^\prime\,-\,
\kappa^2 \dl \stackrel{(2)}{{\bsi}}{}^\prime\,=\,
2\mu \stackrel{(2)}{{\bf S}}\,,
\eqno(3.15)
$$
where prime at $\stackrel{(2)}{{\bsi}}$ implies that the tensor 
is expressed through $\stackrel{(2)}{{\bchi}}{}^\prime$. Provided a 
tensor-valued Green's function of the corresponding operator
acting in L.H.S. of (3.15) is known, solution of (3.15) can be
obtained in a standard way. However, in what follows we shall be 
concerned with (3.14) itself. Therefore, the main task below is to 
adjust (3.14) to the special case in question, i.e., to the case
of the screw dislocation along a cylindric body of circular 
cross-section.

\subsection{The gauge equations in the first and second 
orders. The choice of the model}

Owing to the fact that the equilibrium equations given by (3.11) 
and (3.12) (the first ones in pairs) are fulfilled by  
the Kr\"oner ansatz (3.13), the problem is to determine the 
stress function components $\stackrel{(i)}{{\chi}}_{a b}$ from 
the gauge equations given by (3.11) and (3.12) (the second ones in 
pairs).

Now we replace all the derivatives ${\nb}_a$ by the partial
derivatives $\cd_a\equiv\cd/\cd x^a$, where $x^a$ are
the Cartesian coordinates in the final state \cite{pfl}, 
\cite{gair}. We assume also $\cd_z\equiv\cd_3=0$. Let us introduce 
the following notations for those components of the stress 
potential which are non-trivial:
$$
\begin{array}{l}
\displaystyle{
\mu\stackrel{(i)}{\phi}\,\equiv\,
\cd_2 \stackrel{(i)}{{\chi}}_{1 3}\,-\,
      \cd_1 \stackrel{(i)}{{\chi}}_{2 3}}\,,\qquad\qquad
      i=1,\,2\,,\\ [0.4cm]
f\,\equiv\,\stackrel{(2)}{{\chi}}_{3 3}\,,\quad
p\,\equiv\,-\cd^2_{1 1} \stackrel{(2)}{{\chi}}_{2 2}\,-\,
\cd^2_{2 2} \stackrel{(2)}{{\chi}}_{1 1}\,+\,
2\,\cd^2_{1 2} \stackrel{(2)}{{\chi}}_{1 2}\,.
\end{array}
\eqno(3.16)
$$
The other components $\stackrel{(i)}{{\chi}}_{a b}$ are zero.
The background stress tensor ${\bsi}_{{\rm bg}}$ (see (2.17)) 
is also given by (3.13), though the corresponding stress 
functions are labeled by appropriate subscript: 
$\stackrel{(i)}{{\bchi}}_{{\rm bg}}$.

\vskip 0.7cm
\centerline{{\sl The first order}}
\vskip 0.7cm

Since the background stress field is assumed to correspond to 
that of the screw dislocation in the form provided by
\cite{pfl}, we conclude that only the component of the stress 
potential $\stackrel{(1)}{\phi}$ is nonzero in the first order.
It is described by second equation in (3.11), and the latter
acquires the following form \cite{mal}:
$$
\cd_a\biggl(
\Dl\stackrel{(1)}{\phi}\,-\,
\kappa^2\Bigl(\stackrel{(1)}{\phi}
  \,-\,  \stackrel{(1)}{\phi}_{{\rm bg}}\Bigr)\biggr)
    \,=\,0\,,\qquad a=1,\,2\,,
$$
or
$$
\Dl\stackrel{(1)}{\phi}\,=\,\kappa^2\Bigl(\stackrel{(1)}{\phi}
\,-\,\stackrel{(1)}{\phi}_{{\rm bg}}\Bigr)\,,
\eqno(3.17)
$$
where $\stackrel{(1)}{\phi}_{{\rm bg}}\equiv (-b/2\pi)
\log\rho$. It is appropriate to re-express (3.17) as follows:
$$
\Bigl(\Dl\,-\,\kappa^2\Bigr)\,\Bigl(\stackrel{(1)}{\phi}
  \,-\,\stackrel{(1)}{\phi}_{{\rm bg}}\Bigr)
  \,=\,b\,\dl^{(2)}(x)\,.
\eqno(3.18)
$$

Solution to (3.18) describes the modified screw 
dislocation, and it is given by 
$$
\stackrel{(1)}{\phi}\,=\,\stackrel{(1)}{\phi}_{{\rm bg}}
\,-\,f_S\,,\qquad f_S\,\equiv\,(b/2\pi) K_0 (\kappa\rho)\,.
\eqno(3.19.1)
$$
From (3.13) and (3.16) we obtain the only non-trivial component
of the total stress as follows:
$$
\si_{\phi z}\,=\,
-\,\cd_\rho\Bigl(\mu \stackrel{(1)}{\phi}\Bigr)\,=\,
\frac{b \mu}{2\pi}\,\rho^{\1}
\bigl(1-\kappa\rho K_1(\kappa\rho)\bigr)\,.
\eqno(3.19.2)
$$
Equations (3.19) witness about existence of a core region at 
$\rho\,{\stackrel{<}{_\sim}}\,1/\kappa\,$: outside this region the 
gauge correction to the classical long-ranged law $1/\rho$ is 
exponentially small. At $\rho\ll {\kappa}^{-1}$, the law $1/\rho$ 
(characterizing the stresses) is replaced by another non-singular one.
More detailed information concerning the numerical behaviour of
the solution (3.19.2) (including a treatment of $\kappa^{-1}$
in terms of interatomic spacing) can be found in \cite{laz2}
and \cite{ed3}.

\vskip 0.7cm
\centerline{{\sl The second order}}
\vskip 0.7cm

In order to obtain the gauge equations in this case,
it is necessary to specialize the source $\stackrel{(2)}{S}_{a b}$ 
in (3.14), i.e., its terms $\stackrel{(1)}{Q}{}_{(a b)}$ and 
$\Bigl({\Inc}\stackrel{(1)}{{\bPsi}}\Bigr)_{(a b)}$, which
depend on the first order solution (3.19.1), have to be
written explicitly. (In what follows we shall use
${\bf Q}$, ${\bPsi}$, ${\bf e}$ without the superscript 
$^{(1)}$.)

Let us begin with $Q_{(a b)}$ (3.4.3). More explanations can
be found in \cite{pfl}, \cite{gair}. Using (2.11) we obtain 
from (3.4.3) the following non-zero contributions:
$$
\begin{array}{l}
Q_{1 1}\,=\,Q_{2 2}\,=\,\bigl(\cd_1 e_{2 3}\,-\,
                      \cd_2 e_{1 3}\bigr)^2\,,\\ [0.5cm]
Q_{3 3}\,=\,4\,\bigl(\cd_1 e_{2 3}\,\cd_2 e_{1 3}\,-\,
                      \cd_2 e_{2 3}\,\cd_1 e_{1 3}\bigr)
\,+\,\bigl(\cd_1 e_{2 3}\,-\,\cd_2 e_{1 3}\bigr)^2\,,
\end{array}
\eqno(3.20)
$$
while $Q_{1 2}$, $Q_{2 3}$, $Q_{1 3}$ are zero. Our expressions
(3.20) differ from the analogous quantities in \cite{pfl}, \cite{gair} 
(a different numerical factor in $Q_{1 1}$, $Q_{2 2}$, and absence
of the second term in $Q_{3 3}$) since in our case the corresponding 
Einstein tensor in L.H.S. of the gauge equation (2.15) is expressed 
by means of the Riemann--Christoffel tensor only \cite{mal}. However, in 
\cite{pfl}, \cite{gair} the corresponding incompatibility equation 
includes a contribution from the contortion tensor also. Now let us 
obtain ${\bPsi}$ (3.7) in components:
$$
\begin{array}{l}
\displaystyle{
(1/\mu^2) \Psi_{1 1}\,=\,
-C_3\left[\Bigl(\cd_1\stackrel{(1)}{\phi}\Bigr)^2\,+\,
\Bigl(\cd_2\stackrel{(1)}{\phi}\Bigr)^2\right]\,-\,
C_7 \Bigl(\cd_1\stackrel{(1)}{\phi}\Bigr)^2}\,,\\[0.5cm]
\displaystyle{
(1/\mu^2) \Psi_{2 2}\,=\,
-C_3\left[\Bigl(\cd_1\stackrel{(1)}{\phi}\Bigr)^2\,+\,
\Bigl(\cd_2\stackrel{(1)}{\phi}\Bigr)^2\right]\,-\,
C_7 \Bigl(\cd_2\stackrel{(1)}{\phi}\Bigr)^2}\,,\\[0.5cm]
\displaystyle{
(1/\mu^2) \Psi_{1 2}\,=\,(1/\mu^2) \Psi_{2 1}\,=\,
-C_7\,\cd_1\stackrel{(1)}{\phi} \cd_2\stackrel{(1)}{\phi}}\,,
\\[0.5cm]
\displaystyle{
(1/\mu^2) \Psi_{3 3}\,=\,
-C_3\left[\Bigl(\cd_1\stackrel{(1)}{\phi}\Bigr)^2\,+\,
\Bigl(\cd_2\stackrel{(1)}{\phi}\Bigr)^2\right]
}\,.\\[0.5cm]
\end{array}
\eqno(3.21)
$$

Now we are ready to consider the gauge equations of second order
given by (3.14). We obtain:
$$
\begin{array}{rcl}
\displaystyle{
-\cd^2_{2 2}\Bigl[(1-a)p\,+\,a \Dl f\,-\,
      \kappa^2(f-f_{{\rm bg}})\Bigr]}\,&=&\,2\mu
          \Bigl[Q_{1 1}\,+\,\cd^2_{2 2}\Psi_{3 3}\Bigr]\,,
          \\[0.5cm]
\displaystyle{
-\cd^2_{1 1}\Bigl[(1-a)p\,+\,a \Dl f\,-\,
      \kappa^2(f-f_{{\rm bg}})\Bigr]}\,&=&\,2\mu
          \Bigl[Q_{2 2}\,+\,\cd^2_{1 1}\Psi_{3 3}\Bigr]\,,
          \\[0.5cm]
\displaystyle{
\cd^2_{1 2}\Bigl[(1-a)p\,+\,a \Dl f\,-\,
      \kappa^2(f-f_{{\rm bg}})\Bigr]}\,&=&\,-2\mu\,
          \cd^2_{1 2}\Psi_{3 3}\,,
\end{array}
\eqno(3.22.1)
$$
$$
\begin{array}{rcl}
\displaystyle{
(1-a)\Dl\Dl f\,+\,a \Dl p}&-&
      \kappa^2(p-p_{{\rm bg}})\,=\\[0.5cm] &=&2\mu
          \Bigl[Q_{3 3}\,-\,\cd^2_{1 2}
\bigl(\Psi_{1 2}\,+\,\Psi_{2 1}\bigr)\,+\,\cd^2_{1 1}\Psi_{2 2}
\,+\,\cd^2_{2 2}\Psi_{1 1}\Bigr]\,.
\end{array}
\eqno(3.22.2)
$$
Besides, there exists a couple of equations to determine
$\stackrel{(2)}{\phi}$. But since $\Psi_{1 3}$, $\Psi_{3 1}$,
$\Psi_{2 3}$, $\Psi_{3 2}$ are zero, and 
$\stackrel{(2)}{\phi}_{{\rm bg}}$ is also zero \cite{pfl},
we put consistently $\stackrel{(2)}{\phi}\equiv 0$.
Thus, the only equations to be considered are given by (3.22),
where the corresponding components of ${\bf Q}$ and
${\bPsi}$ are given by (3.20) and (3.21), respectively.

Equations (3.22) look very similiar to those of the classical 
approach \cite{pfl}, \cite{gair} excepting of the fact that 
$Q_{1 1}$, $Q_{2 2}$ are zero in the classical consideration.
The point here is as follows: equations for the classical
stress potentials of second order, $p_{{\rm bg}}$ and 
$f_{{\rm bg}}$, are considered for a two-dimensional domain 
$\rho_c\le\rho\le\rho_e$ ($\rho_e$ and $\rho_c$ are the external 
and internal radii, accordingly). However, $Q_{1 1}$ and $Q_{2 2}$ 
are equal to $4\bigl(\cd_1 e_{2 3}-\cd_2 e_{1 3}\bigr)^2$ in 
\cite{pfl}, and therefore they are proportional to 
$\Dl\stackrel{(1)}{\phi}_{{\rm bg}}$, while the latter represents 
the dislocation density profile (i.e., the dislocation density 
component ${\cal T}^3_{\,\,\,,1 2}$). Outside the disc given by
$0\le\rho\le\rho_c$, $\Dl\stackrel{(1)}{\phi}_{{\rm bg}}$ is zero 
because of its proportionality to the Dirac $\dl$-function. It is 
just the case, why $Q_{1 1}$ and $Q_{2 2}$ drop out of the 
classical version of Eqs. (3.22).

Accordingly to \cite{pfl}, a constant $Q^\prime$ can be introduced
as follows:
$$
Q_{1 1}\,=\,\cd^2_{2 2} Q^{\prime}\,,\quad
Q_{2 2}\,=\,\cd^2_{1 1} Q^{\prime}\,,\quad
Q_{1 2}\,=\,-\cd^2_{1 2} Q^{\prime}\,,
\eqno(3.23)
$$
where $Q^\prime$ is to be adjusted in the end of calculation
of the stress components 
\footnote{More precisely, the numerical value for $Q^\prime$ 
arises from a condition that a mean value of 
$\stackrel{(2)}{\si}_{3 3}$ is zero \cite{pfl}, \cite{gair}.}.
Therefore, equations which define $p_{{\rm bg}}$ and
$f_{{\rm bg}}$ take in our notations the form (compare, for 
instance, with (3.22)):
$$
\begin{array}{lcr}
\displaystyle{
\Dl\Dl f_{{\rm bg}}\,=\,2\mu\,
 \frac a{1-2a}\,\Dl \bigl(\Psi^{{\rm bg}}_{3 3}+Q^\prime \bigr)
 \,+\,2\mu\,\frac{1-a}{1-2a}\,\Bigl[Q^{{\rm bg}}_{3 3}\,+\,
\cd^2_{1 1}\Psi^{{\rm bg}}_{2 2}\,+\,
\cd^2_{2 2}\Psi^{{\rm bg}}_{1 1}\,-\,
2 \cd^2_{1 2} \Psi^{{\rm bg}}_{1 2}\Bigr]}\,,\\[0.5cm]
(1-a)\,p_{{\rm bg}}\,+\,a\,\Dl f_{{\rm bg}}\,=\,
-2\mu\,\bigl(\Psi^{{\rm bg}}_{3 3}+Q^\prime\bigr)\,. 
\end{array}
\eqno(3.24)
$$
The first and the second equations in (3.24) correspond, respectively,
to Eqs. (657) and (655) in \cite{gair}, provided
$\stackrel{(2)}{\si}_{3 3}$ and $F_{(2)}$ therein are identified
as $p_{{\rm bg}}$ and $-f_{{\rm bg}}$. Besides, Eq. (3.23) defines
$Q^\prime$ in opposite way (the sign is different) in comparison with
Eq. (658) in \cite{gair}.

In the present paper, $Q_{1 1}$, $Q_{2 2}$, $Q_{3 3}$ 
are given by (3.20), and we find:
$$
Q_{1 1}\,=\,Q_{2 2}\,=\,
\biggl(\frac12 \Dl\stackrel{(1)}{\phi}\biggr)^2.
\eqno(3.25)
$$
Is it possible to find a ``potential'' $\w g\equiv\w g(\rho)$
which is analogous to $Q^\prime$ of the classical approach?
For such $\w g\equiv\w g(\rho)$ the following equations 
should be respected:
$$
\begin{array}{l}
\displaystyle{
Q_{1 1}\,=\,\cd^2_{2 2}{\w g}\,=\,
\frac12 \Bigl(\cd^2_{\rho \rho}\,
   +\,\frac1{\rho}\cd_\rho\Bigr)\,{\w g}\,-\,
\frac{\cos 2\varphi}2 \,\Bigl(\cd^2_{\rho \rho}\,
   -\,\frac1{\rho}\cd_\rho\Bigr){\w g}}\,,  \\[0.5cm]
\displaystyle{
Q_{2 2}\,=\,\cd^2_{1 1}{\w g}\,=\,
\frac12 \Bigl(\cd^2_{\rho \rho}\,
   +\,\frac1{\rho}\cd_\rho\Bigr) \,{\w g}\,+\,
\frac{\cos 2\varphi}2 \,\Bigl(\cd^2_{\rho \rho}\,
   -\,\frac1{\rho}\cd_\rho\Bigr){\w g}}\,,  \\[0.5cm]
\displaystyle{
Q_{1 2}\,=\,-\cd^2_{1 2}{\w g}\,=\,
-\,
\frac{\sin 2\varphi}2 \,\Bigl(\cd^2_{\rho \rho}\,
   -\,\frac1{\rho}\cd_\rho\Bigr){\w g}}\,=0\,.
\end{array}
\eqno(3.26)
$$
Equations (3.26) lead us to the following pair of
equations:
$$
\begin{array}{l}
\displaystyle{
\Bigl(\cd^2_{\rho \rho}\,-\,\frac1{\rho}\cd_\rho\Bigr){\w g}
\,=\,0}\,,\\[0.5cm]
\displaystyle{
\Bigl(\cd^2_{\rho \rho}\,+\,\frac1{\rho}\cd_\rho\Bigr){\w g}
\,=\,2 Q_{1 1}\,=\,2 Q_{2 2}}\,.
\end{array}
\eqno(3.27)
$$
However, Eqs.(3.27) are not consistent for the given
$Q_{1 1}$, $Q_{2 2}$ (3.25).

In the classical approach, $\Dl\stackrel{(1)}{\phi}_{{\rm bg}}$ 
corresponds to the defect's density profile, and the latter is 
given by the Dirac $\dl$-function. Therefore, $Q_{1 1}=Q_{2 2}=0$ at 
$\rho>\rho_c$, and so it is possible to choose $Q^\prime$
as a constant. In our approach, $\Dl\stackrel{(1)}{\phi}$ also 
represents the density profile of the modified defect, and it 
is given as follows \cite{mal}, \cite{laz2}, \cite{ed3}:
$$
\Bigl(\Dl\stackrel{(1)}{\phi}\Bigr)^2\,=\,
\biggl(\frac{b}{2\pi}\,\kappa^2\,K_0(\kappa\rho)\biggr)^2\,.
\eqno(3.28)
$$
In the limit $\kappa\to\infty$, the function
$\kappa^2\,K_0(\kappa\rho)$ demonstrates a behaviour of
a $\dl$-like function on a plane ``centered'' at $\rho=0$. 
Therefore, the following simplification can seemingly be made 
to keep the 
situation tractable in the framework of the plane problem: let us 
approximate $\kappa^2\,K_0(\kappa\rho)$ by a piecewisely constant 
function which takes two different constant values either within 
the disc $0\le\rho\le\rho_c$, or outside it. Besides, we shall 
assume that differentiations of this ``hat''-function at 
$\rho=\rho_c\pm 0$ are negligible. Then, it turns out 
that equations (3.27) become consistent. Clearly, such 
approximation is rather rough at $\rho\ll \kappa^{-1}$, i.e., in a 
very close vicinity of the classical defect's axis. However, as we 
shall see below, this simplification leads to a reasonable 
picture for a non-singular modified screw dislocation in the
second order also.

The replacement proposed for the density profile is given as:
$$
\frac{b}{2\pi}\,\kappa^2\,K_0(\kappa\rho)\,\longmapsto\,
\frac{b}{\pi \rho^2_c}\,h_{[0,\,\rho_c]}(\rho)\,,
\eqno(3.29)
$$
where $h_{[0,\,\rho_c]}(\rho)$ is equal to unity at
$0\le\rho\le\rho_c$, and to zero otherwise. 
With the density profile given by (3.29), we find
${\w g}$  which respects (3.27):
$$
{\w g}(\rho)\,=\,\left[
\begin{array}{llr}
{\cC}^\prime\,,\qquad\qquad &\rho_c <\rho\,,&\\[0.4cm]
\displaystyle{
\biggl(\frac{b}{2\pi\rho^2_c}\biggr)^2\,(\rho^2/2)}
\,+\,{\cC}^{\prime \prime}\,, &0\le\rho\le\rho_c\,,&
\end{array}\right.
\eqno(3.30)
$$
where the constants ${\cC}^\prime$ and ${\cC}^{\prime \prime}$
will be adjusted later. 

Let us write $Q_{3 3}$ (3.20) explicitly:
$$
Q_{3 3}\,=\,\Bigl(\cd^2_{1 2}\stackrel{(1)}{\phi}\Bigr)^2
\,-\,\cd^2_{1 1}\stackrel{(1)}{\phi}\,
\cd^2_{2 2}\stackrel{(1)}{\phi}\,+\,
\biggl(\frac12 \Dl\stackrel{(1)}{\phi}\biggr)^2\,.
\eqno(3.31)
$$
Using (3.21) and (3.31), we calculate:
$$
\begin{array}{l}
Q_{3 3}\,-\,2 \cd^2_{1 2} \Psi_{1 2}\,+\,\cd^2_{1 1}\Psi_{2 2}
\,+\,\cd^2_{2 2}\Psi_{1 1}\,=\\[0.5cm]
\,=\,
\displaystyle{
\Dl\Psi_{3 3}\,+\,
\bigl(1-2\mu^2 C_7\bigr)\left[
\Bigl(\cd^2_{1 2}\stackrel{(1)}{\phi}\Bigr)^2
\,-\,\cd^2_{1 1}\stackrel{(1)}{\phi}\,
\cd^2_{2 2}\stackrel{(1)}{\phi}\right]\,+\,
\biggl(\frac12 \Dl\stackrel{(1)}{\phi}\biggr)^2}\,.
\end{array}
\eqno(3.32)
$$
Therefore, taking into account (3.26) and (3.32), we obtain 
from (3.22) the following couple of equations to determine
$f$ and $p$:
$$
(1-a)p\,+\,a \Dl f\,-\,\kappa^2(f-f_{{\rm bg}})\,=\,
-2\mu\,(\Psi_{3 3}\,+\,{\w g})\,,
\eqno(3.33)
$$
$$
(1-a)\Dl\Dl f\,+\,a \Dl p
\,-\,\kappa^2(p-p_{{\rm bg}})\,=\,
\frac{\mu}2\,\Bigl(\Dl\stackrel{(1)}{\phi}\Bigr)^2\,+
$$
$$+\,
2\mu\,\Dl\Psi_{3 3}\,+\,2\mu
\,\bigl(1-2\mu^2 C_7\bigr)\left[
\Bigl(\cd^2_{1 2}\stackrel{(1)}{\phi}\Bigr)^2
\,-\,\cd^2_{1 1}\stackrel{(1)}{\phi}\,
\cd^2_{2 2}\stackrel{(1)}{\phi}\right]
\,.
\eqno(3.34)
$$
We obtain $p$ from (3.33) and substitute it into (3.34):
$$
\displaystyle{
p\,=\,-\frac{a}{1-a}\,\Dl f\,+\,\frac{\kappa^2}{1-a}
\,(f-f_{{\rm bg}})\,-\,
\frac{2\mu}{1-a}\,(\Psi_{3 3}\,+\,{\w g})}\,,
\eqno(3.35)
$$
$$\displaystyle{
\bigl(\Dl\,-\,\kappa^2\bigr)\,\Bigl(\Dl\,+\,
\frac{\kappa^2}{1-2a}\Bigr)\,(f-f_{{\rm bg}})\,=\,{\cR}}\,,
\eqno(3.36)
$$
where
$$
\begin{array}{rcl}
\displaystyle{
\frac{1-2a}{2\mu}}\,{\cR}\,&\equiv&\,
\bigl(\Dl\,-\,\kappa^2\bigr)
\bigl(\Psi_{3 3}\,+\,{\w g}\,-\,\Psi^{{\rm bg}}_{3 3}
\,-\,Q^\prime\bigr)\,+\, \\[0.5cm]
&+&\,
\displaystyle{
(1-a)\bigl(1\,-\,2\mu^2 C_7\bigr)
\bigl(\Phi\,-\,\Phi_{{\rm bg}}\bigr)\,-\,\frac{1-a}4
\Bigl(\Dl\stackrel{(1)}{\phi}\Bigr)^2}\,,
\end{array}
\eqno(3.37)
$$
$$
\Phi\,=\,\Bigl(\cd^2_{1 2}\stackrel{(1)}{\phi}\Bigr)^2
\,-\,\cd^2_{1 1}\stackrel{(1)}{\phi}\,
\cd^2_{2 2}\stackrel{(1)}{\phi}\,.
\eqno(3.38)
$$
Here $f_{{\rm bg}}$ respects the first equation in (3.24),
and $\Phi_{{\rm bg}}$, $\Psi^{{\rm bg}}_{3 3}$ are
given by (3.38), (3.21) (provided $\stackrel{(1)}{\phi}$ is 
replaced by $\stackrel{(1)}{\phi}_{{\rm bg}}$), correspondingly.
Besides, we formally keep $(\Dl\stackrel{(1)}{\phi})^2$ 
as an exact expression.

\subsection{Final remarks about the gauge equations}

To conclude Section 3, we should pay attention also
to the structure of the function ${\cR}$ in the righ-hand 
side of (3.36). Explicit expression for ${\cR}$ is given
by Eq. (3.37). Our problem has an axial symmetry, and therefore 
the use of the cylindric coordinates $\rho, \ph, z$ 
instead of the coordinates in final state $x^a, a=1, 2, 3$
(the coordinates $\rho$ and $\ph$ are chosen in 
$(x^1, x^2)$-plane and $z\equiv x_3$) is more appropriate. 
Therefore, we obtain from (3.21) the following expressions
in the cylindrical coordinates:
$$
\begin{array}{l}
\displaystyle{
\Psi_{3 3}\,=\,-c\,\bigl(\cd_\rho\stackrel{(1)}{\phi}\bigr)^2\,,
\qquad c\,\equiv\,\mu^2 C_3}\,,\\[0.5cm]
\Dl \Psi_{3 3}\,=\,-2 c\left[
\cd_\rho\stackrel{(1)}{\phi}\,
\cd_\rho\bigl(\Dl \stackrel{(1)}{\phi}\bigr)\,+\,
\bigl(\Dl \stackrel{(1)}{\phi}\bigr)^2\right]\,-\,
4 c\,\Phi\,,
\end{array}
\eqno(3.39)
$$
where $\Phi$ (3.38) is re-written as:
$$
\Phi\,=\,-\frac 1{\rho}\,\cd_\rho\stackrel{(1)}{\phi}
\cd^2_{\rho \rho}\stackrel{(1)}{\phi}\,=\,
-\frac 1{\rho}\,\cd_\rho\stackrel{(1)}{\phi}
\Dl \stackrel{(1)}{\phi}\,+\,
\frac{(\cd_\rho\stackrel{(1)}{\phi})^2}{\rho^2}\,.
\eqno(3.40)
$$
Further, we use (3.39), (3.40) in (3.37), and obtain
${\cR}$ in the following form:
$$
\begin{array}{rcl}
{\cR}\,&=&\,
\displaystyle{k\left[W(\rho)\,+\,
\biggl(\frac{{\tilde c}}{\rho^2}\,+\,\kappa^2 c\biggr)\,\biggl(
\Bigl(\cd_\rho\stackrel{(1)}{\phi}\Bigr)^2\,-\,
\Bigl(\cd_\rho\stackrel{(1)}{\phi}_{{\rm bg}}\Bigr)^2\biggr)
\right.  }\\[0.5cm]
&-&\displaystyle{
\frac{{\tilde c}}{\rho}\Bigl(
\Dl \stackrel{(1)}{\phi} \cd_\rho\stackrel{(1)}{\phi}\,-\,
\Dl \stackrel{(1)}{\phi}_{{\rm bg}}
\cd_\rho\stackrel{(1)}{\phi}_{{\rm bg}}\Bigr)
}\\[0.5cm]
&-&\displaystyle{\left.
2c\biggl(\cd_\rho\Bigl(
\Dl \stackrel{(1)}{\phi}\Bigr) \cd_\rho\stackrel{(1)}{\phi}\,-\,
\cd_\rho\Bigl(\Dl \stackrel{(1)}{\phi}_{{\rm bg}}\Bigr)
\cd_\rho\stackrel{(1)}{\phi}_{{\rm bg}}\biggr)        \right]
}  \,,
\end{array}
\eqno(3.41)
$$
where
$$
\begin{array}{l}
\displaystyle{
k\,\equiv\,\frac{2\mu}{1-2a}\,,\qquad {\tilde c}\,\equiv\,
(1-a)(1-2\mu^2 C_7)\,-\,4c
}\,,\\[0.5cm]
\displaystyle{
W(\rho)\,\equiv\,\frac{1+a}4\,
\Bigl(\Dl \stackrel{(1)}{\phi}\Bigr)^2\,-\,
2 c\biggl( \Bigl(\Dl \stackrel{(1)}{\phi}\Bigr)^2\,-\,
\Bigl(\Dl \stackrel{(1)}{\phi}_{{\rm bg}}\Bigr)^2\biggr)
\,-\,\kappa^2 \bigl({\w g}\,-\,Q^\prime\bigr)}\,.
\end{array}
\eqno(3.42)
$$

Let us have a look at ${\cR}$ (3.41). The contribution most 
interesting for us is given by the second term in it. The other 
terms in (3.41) contain either $\Dl \stackrel{(1)}{\phi}$ or 
$\cd_\rho\Bigl(\Dl \stackrel{(1)}{\phi}\Bigr)$, i.e., 
are dependent on the density profile or on its derivatives. In 
other words, these terms are more significant either within the
core or near its boundary. 

Let us rewrite ${\cR}$ (3.41) once again using
the explicit expression for the background solution of the first 
order $\stackrel{(1)}{\phi}_{{\rm bg}}\equiv (-b/2\pi)\log\rho$. 
We obtain:
$$
\begin{array}{rcl}
{\cR}\,&=&\,
\displaystyle{
k\left[ W(\rho)\,+\,
\biggl(\frac{{\tilde c}}{\rho^2}\,+\,\kappa^2 c\biggr)\,
\cd_\rho f_S\,\Bigl(\cd_\rho f_S\,-\,2 \cd_\rho
\stackrel{(1)}{\phi}_{{\rm bg}}\Bigr)
\right.  }\\[0.5cm]
&+&\displaystyle{
{\tilde c}\,\Dl \stackrel{(1)}{\phi} \frac{\cd_\rho f_S}{\rho}\,+\,
2 c\,\cd_\rho\Bigl(\Dl \stackrel{(1)}{\phi}\Bigr)\cd_\rho f_S
}\\[0.5cm]
&-&\displaystyle{
\Bigg.
{\tilde c}\,\Bigl(\Dl \stackrel{(1)}{\phi}\,-\,
\Dl \stackrel{(1)}{\phi}_{{\rm bg}}\Bigr)
\frac{\cd_\rho\stackrel{(1)}{\phi}_{{\rm bg}}}{\rho}
-\,2c\,\cd_\rho\Bigl(\Dl \stackrel{(1)}{\phi}\,-\,
\Dl \stackrel{(1)}{\phi}_{{\rm bg}}\Bigr)
\cd_\rho\stackrel{(1)}{\phi}_{{\rm bg}}\Bigg]
}  \,.
\end{array}
\eqno(3.43)
$$
Now the structure of ${\cR}$ can be characterized as follows. The first 
term, $W(\rho)$, is determined by the density profile, and it seems to 
be significant rather within the core since the constant $Q^{\prime}$ 
can be removed outside $0\le \rho\le \rho_c$ by an appropriate choice 
of ${\cC}^{\prime}$ in ${\w g}$ (3.30). The second term in (3.43) 
behaves as $\rho^{-4}$ at $\rho\ll 1$, and therefore it is just 
responsible for the fact that the stress function to be found $f$ is 
expected to cancel exactly the most important term ($\propto\log^2\rho$) 
in $f_{\rm bg}$ at $\kappa\rho\ll 1$:
$$
\Dl\Dl (f\,-\,f_{\rm bg})\,=\,- 
\displaystyle{\frac{b^2}{4\pi^2}\frac{k {\tilde c}}{\rho^4}}\,,
\quad
\Dl\Dl f_{\rm bg}\,=\,
\displaystyle{\frac{b^2}{4\pi^2}\frac{k {\tilde c}}{\rho^4}}\,,
$$
while $f_{\rm bg}$ itself is given by Eq. (5.12) below. The third and 
the fourth terms are concerned with the density profile and with its 
derivatives. It can be assumed that the last two terms imply an 
increment of the corresponding quantities in the gauge approach 
due to a replacement of the density profile of the background defect 
by the density profile of the modified defect. In what follows, we shall 
neglect a possible influence of the last two contributions. Thus we obtain:
$$
\begin{array}{rcl}
{\cR}\,\approx\,k\Bigg[ W(\rho)\,&+&\,
\displaystyle{
\biggl(\frac{{\tilde c}}{\rho^2}\,+\,\kappa^2 c\biggr)\,
\cd_\rho f_S\,\Bigl(\cd_\rho f_S\,-\,2 \cd_\rho
\stackrel{(1)}{\phi}_{{\rm bg}}\Bigr)
}
\Bigg. \\[0.5cm]
&+&\displaystyle{\Bigg.
{\tilde c}\,\Dl \stackrel{(1)}{\phi} \frac{\cd_\rho f_S}{\rho}\,+\,
2 c\,\cd_\rho\Bigl(\Dl \stackrel{(1)}{\phi}\Bigr)\cd_\rho f_S
\Bigg]}\,,
\end{array}
\eqno(3.44)
$$
where we put approximately
$$
W(\rho)\,\approx\,\frac{1+a}4\,
\Bigl(\Dl \stackrel{(1)}{\phi}\Bigr)^2\,-\,
\kappa^2 \bigl({\w g}\,-\,Q^\prime\bigr)\,.
\eqno(3.45)
$$

\section{Solution of the gauge equation}

\subsection{Preparation}

Now the task is to solve Eq. (3.36) with the R.H.S. given by 
${\cR}$ (3.44) with $W(\rho)$ (3.45). We shall do it in two steps. 
As the first step we shall solve equation
$$
\left[z^2\frac{d^2}{d z^2}\,+
\,z \frac{d}{d z}\,-\,z^2\right] G(z)\,=\,k\,{\cR}(z)\,,
\eqno(4.1)
$$
where the variable $z$ implies the radial coordinate $\rho$ 
rescaled as follows: $z=\kappa\rho$, $d/dz=\kappa^{\1} d/d \rho$. 
Equation (4.1) is a non-homogeneous Bessel equation \cite{mag}, and 
its new source (after multiplication of equation by $\rho^2$),
though again denoted by ${\cR}$, is written in the new variables:
$$
\begin{array}{rcl}
{\cR}(z)\,&\equiv&\,\displaystyle{
w_c(z) z^2\,+\,X^2\bigl({\tilde c}\,+\,c z^2\bigr)\,K_1(z)\,
\Bigl(K_1(z)\,-\,\frac 2z\Bigr)}
\\[0.5cm]
&-& 
\displaystyle{
X^2 z\,K_1(z) \left({\tilde c}\,\Dl_z {\w\phi}\,+\, 2 c\,z 
\frac{d}{d z}\Bigl(\Dl_z {\w\phi}\Bigr)\right)\,,
}
\end{array}
\eqno(4.2)
$$
where $X^2\equiv\kappa^2\displaystyle{\frac{b^2}{4\pi^2}}$,
$\Dl_z\equiv\displaystyle{\frac{d^2}{d z^2}+
z^{\1} \frac{d}{d z}}$, $K_1(z)$ is the modified 
Bessel function \cite{mag}, and ${\w\phi}$ implies 
$\displaystyle{\stackrel{(1)}{\phi}}$ with removed
factor $\displaystyle{\frac{b}{2\pi}}$.

In order to explain the notation $w_c(z)$ in (4.2), let us have 
a look at $W(\rho)$ (3.45). Let us choose the constant 
${\cC}^\prime$, which appears in ${\w g}$ (3.30), equal to 
the constant $Q^\prime$ defined by (3.23). Then we obtain for 
$W(\rho)$:
$$
\displaystyle{
W(\rho)\,=\,\biggl(\frac{b}{2\pi}\biggr)^2
\frac1{\rho^4_c}\left[(1\,+\,a)\,-\,\frac{\kappa^2}2
\bigl(\rho^2\,-\,\rho^2_c\bigr)\right]\,h_{[0,\,\rho_c]}(\rho)}\,,
\eqno(4.3)
$$
where ${\cC}^{\prime\prime}$ (see ${\w g}$ (3.30)) is fixed by 
requirement of continuity of the density profile at $\rho_c$:
$$
\displaystyle{{\cC}^{\prime\prime}\,+\,\frac{b^2}{8\pi^2}
\frac1{\rho^2_c}\,=\,Q^{\prime}}\,.
$$
After the replacement of $\kappa\rho$ by $z$, we obtain for
$\kappa^{-2} W(\rho)$:
$$
\displaystyle{
\frac{X^2}{z^4_c}\left[(1+a)\,-\,
      \frac12\bigl(z^2-z^2_c\bigr)\right]\,
           {\w h}_{\,[0,\, z_c]}(z)}\,\equiv\,w_c(z)\,,
\eqno(4.4)
$$
where $z_c\equiv\kappa\rho_c$, and ${\w h}_{\,[0,\, z_c]}(z)$
is equal to $1$ at $z\in [0,\, z_c]$ or to zero, otherwise.

Solution of (4.1) is based on a knowledge of the asymptotical 
behaviour of $\cR$ (4.2) at $z\gg 1$ and $z\ll 1$. Let us 
obtain the corresponding expansions. First of all, 
let us take into account that $\Dl_z {\w\phi}$ is exponentially 
small at large $z$ (or even zero, provided the replacement (3.29) 
is made) since it is equal to $-K_0(z)$. Therefore, 
$\cR$ is localized at large $z$.

In order to study the case $z\ll 1$, let us represent
$\cR$ as a sum: 
$$
{\cR}\equiv w_c(z) z^2+{\cR}_1(z)+{\cR}_2(z)+{\cR}_3(z)\,.
$$
We obtain the following expansions for 
${\cR}_1$, ${\cR}_2$, ${\cR}_3$ at small $z$:
$$
\begin{array}{rcl}
\!\!\!\!\!\displaystyle{\!\!\!\!\!
\frac{{\cR}_1}{X^2}}
\,\equiv\!
&\bigl(&\displaystyle{\!\!\!{\tilde c}\,+\,
c z^2\bigr)\,K_1(z)\,\Bigl(K_1(z)\,-\,\frac 2z\Bigr)}\\[0.5cm]
\simeq\! &-&\displaystyle{
\!\!\!\frac{{\tilde c}}{z^2}\,-\,c\,+\,
\frac{{\tilde c}}{4}\,z^2\log^2z\,-\,
\frac{{\tilde c}}4\Bigl(1-2\log\frac \ga{2}\Bigr)\,z^2\log z}\\[0.5cm]
\!&+&\displaystyle{
\!\!\!\frac {\tilde c}{16} \Bigl(1-2\log\frac \ga{2}\Bigr)^2\,z^2\,+\,
o(z^2 \log^2z)}\,;
\end{array}
\eqno(4.5)
$$
$$
\begin{array}{rcl}
\displaystyle{
\frac{{\cR}_2}{X^2}}&\equiv &
\displaystyle{
-{\tilde c}\,z\,K_1(z)\Dl_z {\w\phi}\,\approx\,
\frac{2{\tilde c}}{z^2_c}\,z K_1(z)\, 
{\w h}_{\,[0,\, z_c]}(z)}\\ [0.5cm]
&\simeq&
\displaystyle{\frac{2{\tilde c}}{z^2_c}\,{\w h}_{\,[0,\, z_c]}(z)
\,\left(1\,+\,\frac12 z^2\log z\,-\,
\frac{z^2}{4}\Bigl(1-2\log\frac \ga{2}\Bigr)\,+\,o(z^2)\right)}\,;
\end{array}
\eqno(4.6)
$$
$$
\begin{array}{rcl}
\!\!\!\!\!
\!\!\!\!\!\!\!\!\!\!\!\displaystyle{\frac{{\cR}_3}{X^2}}\,&\equiv &
\displaystyle{
-2 c\,z^2 K_1(z)\frac{d}{d z}\Bigl(\Dl_z {\w\phi}\Bigr)}\\ [0.5cm]
&\simeq &\,
\displaystyle{
-2 c\,-\,2 c\,z^2\log z\,+\,c \Bigl(1-2\log\frac \ga{2}\Bigr)
z^2\,+\,o(z^2)}\,.
\end{array}
\eqno(4.7)
$$
In (4.6), we took into account the replacement (3.29). 
For a comparison, the exact expansion for ${\cR}_2$ looks
as follows:
$$
\displaystyle{\,\,\,\,\,\,\frac{{\cR}_2}{X^2}
\,\simeq\,-{\tilde c}\log\Bigl(\frac \ga{2} z\Bigr)\,+\,
\frac{{\tilde c}}4 z^2
\left(1-2\log^2\Bigl(\frac \ga{2} z\Bigr)\right)}\,+\,
o(z^2)\,.
\eqno(4.8)
$$
It is seen from (4.8) that the terms $\propto\log z$ and 
$\propto z^2\log^2 z$ are absent in the approximate expression (4.6).

Expansions (4.4)--(4.8) suggest the following series form of
$\cR$ (4.2) at $z\ll 1$:
$$
\begin{array}{rcl}
{\cR}(z)&\simeq &p_0\,z^2\,+\,p_1\,z^{-2}\,+\,p_2\,\log z
\,+\,p_3\\[0.5cm]
&+&p_4\,z^2 \log^2 z\,+\,p_5\,z^2 \log z\,+
\,p_6\,z^2\,+\,o(z^2)\,,
\end{array}
\eqno(4.9)
$$
where $p_0\equiv p_0 (z)$ implies $w_c(z)$ (4.4). The coefficients
$p_1$, $p_2$, $p_3$, $p_4$, $p_5$, $p_6$ are influenced by our
assumptions, and we obtain them as follows:
$$
\begin{array}{l}
p_1\,=\,-X^2\,{\tilde c}\,,\\[0.5cm]
\displaystyle{
p_3\,=\,X^2\biggl(\frac{2 {\tilde c}}{z^2_c}\,-
\,c\biggr)}\,,\\[0.5cm]
\displaystyle{
p_4\,=\,X^2\,\frac{{\tilde c}}4}\,,\\[0.5cm]
\displaystyle{
p_5\,=\,X^2\biggl(\frac{{\tilde c}}{z^2_c}\,-\,
\frac{{\tilde c}}4
\Bigl(1-2\log \frac \ga{2}\Bigr)\biggr)}\,,\\[0.5cm]
\displaystyle{
p_6\,=\,X^2 \frac{{\tilde c}}{16}\,\Bigl(1-2\log \frac \ga{2}\Bigr)
\,\Bigl(1-2\log \frac \ga{2}\,-\,\frac 8{z^2_c}\Bigr)}\,.
\end{array}
\eqno(4.10)
$$
Practically, only ${\cR}_1$ (4.5) and ${\cR}_2$ (4.6) are taken into 
account in order to assign the specific values (4.10) to the 
corresponding coefficients in the expansion (4.9). The contribution 
${\cR}_3$ (4.7) is excluded from the consideration
since we use (3.29) and neglect possible contributions which should 
be important near the boundary of the core. However, a possible 
implication of ${\cR}_3$ for the coefficients in (4.9) would 
be given by shifts in $p_3$, $p_5$, $p_6$ by numerical constants 
dependent on $c$ (3.39) (for instance, $-c$ in $p_3$ would be replaced
by $-3c$). A usage of ${\cR}_2$ in the form (4.8) instead of the 
approximate expression (4.6) would remove in $p_3$, $p_5$, $p_6$ (4.10) 
the dependence on the value of the core radius $z_c$. Besides, 
usage of (4.8) would lead to non-zero contributions in $p_2$ and $p_4$.
But since our choice of the density profile is given by 
(3.29), the coeficient $p_2$ is simply zero. By formal reasons, 
we keep the corresponding terms in (4.9) with unspecified $p_2$ which 
is to be equated to zero at the very end of the calculation.

\subsection{The solution }

Therefore, let us first consider the non-homogeneous equation:
$$
\left[z^2\frac{d^2}{d z^2}\,+
\,z \frac{d}{d z}\,-\,z^2\right] y(z)\,=\,k\,{\cR}(z)\,.
\eqno(4.11)
$$
General solution to (4.11) is  given by the standard formula \cite{mag}:
$$
\begin{array}{rcl}
k^{\1}\,y(z) &=& A\,y_1(z)\,+\,B\,y_2(z) \\[0.5cm]
&-&\displaystyle{
\int\limits^z_{z_0}
\frac{y_1(z)\,y_2(t)\,-\,y_2(z)\,y_1(t)}
     {y_1(t)\,y_2^\prime(t)\,-\,y_2(t)\,y_1^\prime(t)}}
\,\frac{{\cR}(t)}{t^2}\,\,dt\,.
\end{array}
\eqno(4.12)
$$
Here $z_0$ is to be adjusted, and $y_1(z)$, $y_2(z)$ 
are linearly independent solutions of the corresponding 
homogeneous equation.

In our case we put $y_1(z)=I_0(z)$ and $y_2(z)=K_0(z)$ \cite{mag},
and (4.12) gives us solution to (4.11) as follows. Let us
define $G(z, s)$:
$$
\begin{array}{rcl}
k^{\1}\,G(z, s)&=&
\displaystyle{
I_0(z)\,\left(A(s)\,+\,
\int\limits^z_s K_0(t) {\cR}(t)\,\frac{dt}{t}\right)}\\
[0.5cm]
&+&\displaystyle{
K_0(z)\,\left(B(s)\,-\,
\int\limits^z_s I_0(t) {\cR}(t)\,\frac{dt}{t}\right)}\,,
\end{array}
\eqno(4.13.1)
$$
where
$$
\begin{array}{l}
A(s)\,\equiv\,-\,\displaystyle{
\int\limits^\infty_s K_0(t) {\cR}(t)\,\frac{dt}{t}}\,,\\ [0.5cm]
B(s)\,\equiv\,const\,+\,\displaystyle{\int\limits^1_s
\left(\frac{p_1}{t^3}\,+\,\frac{p_2}{t}\,\log t\,+\,
\Bigl(\frac{p_1}4\,+\,p_3\Bigr)\frac 1t\right)\,dt}\,.
\end{array}
\eqno(4.13.2)
$$
Then, solution to (4.11) appears as
$$
G(z)\,=\,\lim_{s\to 0}\,G(z, s)\,\equiv\, G(z, 0)\,.
\eqno(4.14)
$$

Asymptotical behaviour of $G(z)$ can be obtained from
(4.13), (4.14). At large $z$, $G(z)$ decays exponentially,
i.e., $G(z)$ is well localized. Using the expansions
provided in Appendix $A$, it is straightforward
to establish the behaviour of $G(z)$ at small $z$:
$$
\begin{array}{rcl}
k^{\1}\,G(z)&\simeq &q_0\,+\,q_1\,z^{-2}\,
+\,q_2\,\log^3 z\,+\,q_3\,\log^2 z\,+\,q_4\,\log z
\\[0.5cm]
&+&q_5\,z^2 \log^3 z\,+\,q_6\,z^2 \log^2 z\,+
\,q_7\,z^2 \log z\,+\,q_8\,z^2\,,
\end{array}
\eqno(4.15)
$$
where
$$
\begin{array}{l}
\displaystyle{
q_0\,=\,-const\times\log\frac{\ga}2\,-\,
{\cal I}_K\,-\,\frac{p_1}{16}}\,,\\[0.5cm]
\displaystyle{
q_1\,=\,\frac{p_1}4\,,\quad q_2\,=\,\frac{p_2}6\,,
\quad q_3\,=\,\frac{p_1}8\,+\,\frac{p_3}2}
\,,\quad q_4\,=\,-\,const\,+\,\frac{3 p_1}4\,,\\[0.5cm]
\displaystyle{
q_5\,=\,\frac{p_2}{24}\,,\quad q_6\,=\,\frac{p_1}{32}
\,+\,\frac{p_3-p_2}8\,+\,\frac{p_4}4\,,}\\[0.5cm]
\displaystyle{
q_7\,=\,-\frac{const}4\,+\,\frac{p_1}{8}\,+\,\frac{3 p_2}{16}
\,-\,\frac{p_3}4\,-\,\frac{p_4}2\,+\,\frac{p_5}4\,,
\quad  q_8\,=\,\frac{p_0\,-\,{\cal I}_K}4\,+}\\[0.5cm]
\displaystyle{
\qquad\,+\,\frac{const}4
\Bigl(1\,-\,\log\frac{\ga}2\Bigr)\,-\,\frac{5 p_1}{32}
\,-\,\frac{p_2}8\,+\,\frac{3 p_3}{16}\,+\,\frac{3 p_4}{8}
\,+\,\frac{p_6\,-\,p_5}4\,.}
\end{array}
\eqno(4.16)
$$
In (4.16), $p_1,\dots, p_6$ are given by (4.10), ${\cal I}_K$ is 
given by (A8) in Appendix A, and ${\it const}$ is introduced by the
definition of $B(s)$ (4.13.2). It is seen that the term $p_2\log z$ 
in (4.9) is just responsible for the highest powers of the 
logarithm in (4.15): $q_2\log^3 z$ and $q_5 z^2\log^3 z$ (and 
similarly for the logarithms at higher powers of $z^2$).

As a second step, we are going to find the modified stress potential
of second order $f\equiv f_{\rm bg}+{\cF}$, where ${\cF}$ respects
the Bessel equation
$$
\left[z^2\frac{d^2}{d z^2}\,+
\,z \frac{d}{d z}\,+\,z^2\right] {\cF}(z)\,=\,
\frac{z^2}{{\cN}^2}\,G\,\Bigl(\frac{\kappa}{{\cN}}\,z\Bigr)\,.
\eqno(4.17)
$$
In (4.17), the variable $z$ is defined differently, $z\equiv {\cN}\rho$, 
${\cN}^2\equiv\displaystyle{\frac{\kappa^2}{1-2a}}$, and $G(z)$ is given 
by (4.13), (4.14). Using again (4.12), we obtain solution to (4.17) in 
the following form:
$$
\begin{array}{l}
{\cF}(\rho)\,=\,C\,{\w Y}_0({\cN}\rho)\,+\,D\,J_0({\cN}\rho)
\,+\,I_{{\cF}}(\rho)\,,\\[0.5cm]
I_{{\cF}}(\rho)\,\equiv\,
\displaystyle{
J_0({\cN}\rho)\,\int\limits^\infty_\rho {\w Y}_0({\cN} t) 
G(\kappa t)\,t\,dt}\,
-\,\displaystyle{{\w Y}_0({\cN}\rho)\,
\int\limits^\infty_\rho J_0({\cN}t) G(\kappa t)\,t\,dt}\,,
\end{array}
\eqno(4.18)
$$
where ${\w Y}_0(z)\equiv (\pi/2)Y_0(z)$, and $Y_0(z)$,
$J_0(z)$ are the Bessel functions which solve the homogeneous 
version of (4.17).

For our purposes it is appropriate to put $D=0$ and $C\ne 0$ in 
(4.18). Now let us write the asymptotical results for ${\cF}(\rho)$. 
At large $\rho$, the contribution given by $I_{{\cF}}$ is 
exponentially small, and thus ${\w Y}_0({\cN}\rho)$ dominates in 
${\cF}(\rho)$ at ${\cN}\rho\gg 1$:
$$
\displaystyle{
C\,{\w Y}_0({\cN}\rho)\,\simeq\,
C\,\biggl(\frac{\pi}{2 {\cN}\rho}\biggr)^{1/2}\,
\!\!\!\!\sin\Bigl({\cN}\rho\,-\,\frac\pi{4}\Bigr)}\,.
$$
For small ${\cN}\rho$, we obtain (Appendix $B$):
$$
\begin{array}{rcl}
{\cF}(\rho)&\simeq &r_0\,+\,r_1\,\log \rho\,
+\,r_2\,\log^2 \rho \\[0.5cm]
&+&r_3\,\rho^2 \log^3 \rho\,+\,r_4\,\rho^2 \log^2 \rho\,+
\,r_5\,\rho^2 \log \rho\,+\,r_6\,\rho^2\,+\,r_7\,\rho^4 
\log^3 \rho\,,
\end{array}
\eqno(4.19)
$$
where
$$
\begin{array}{l}
\displaystyle{
r_0\,=\,- {\w {\cal I}_Y}\,+\,\log\Bigl(\frac{\ga}2{\cN}\Bigr)
\,\bigl(C\,+\,{\w {\cal I}_J}\bigr) }\,,\\[0.5cm]
\displaystyle{
r_1\,=\,C\,+\,{\w {\cal I}_J}\,,\qquad r_2\,=\,k\,
\frac{p_1}{8\kappa^2}\,,\qquad r_3\,=
                \,k\,\frac{p_2}{24}}\,,\\[0.5cm]
\displaystyle{
r_4\,=\,{\h J}_5\,-\,{\h Y}_5\,+\,3\log{\cN}\,
\bigl({\h J}_4\,-\,{\h Y}_4 \bigr)}\\[0.5cm]
\quad \displaystyle{
\,=\,k\,\left[
\biggl(1\,-\,\frac{{\cN}^2}{\kappa^2} \biggr)\,
\frac{p_1}{32}\,-\,(1\,-\,\log\kappa)\frac{p_2}8\,
+\,\frac{p_3}8 \right]\,,}\\[0.5cm]
\displaystyle{
r_5\,=\,{\h J}_6\,-\,{\h Y}_6\,+\,2\log{\cN}\,
\bigl({\h J}_5\,-\,{\h Y}_5 \bigr)\,+\,3\log^2{\cN}\,
\bigl({\h J}_4\,-\,{\h Y}_4 \bigr)\,
             -\,C\,\frac{{\cN}^2}4 }\,,\\[0.5cm]
\displaystyle{
r_6\,=\,{\h J}_7\,-\,{\h Y}_7\,+\,\log{\cN}\,
\bigl({\h J}_6\,-\,{\h Y}_6 \bigr)\,+\,\log^2{\cN}\,
\bigl({\h J}_5\,-\,{\h Y}_5 \bigr)}\\[0.5cm]
\displaystyle{
\quad +\,\log^3{\cN}\,
\bigl({\h J}_4\,-\,{\h Y}_4 \bigr)\,+\,C\,\frac{{\cN}^2}4
\left(1\,-\,\log\Bigl(\frac{\ga}2{\cN}\Bigr)\right) 
}\,,\\[0.5cm]
r_7\,=\,{\h J}_9\,-\,{\h Y}_9\,=\,k 
\displaystyle{\frac{\kappa^2-{\cN}^2}{384}} p_2\,.
\end{array}
\eqno(4.20)
$$
In (4.20), ${\w {\cal I}_Y}$, ${\w {\cal I}_J}$, and
${\h J}_i$, ${\h Y}_i$ ($i=4,\dots, 9$) are defined
in Appendix B, and $p_1,\dots, p_6$ are given by Eqs. (4.10).

\section{The stress tensor}
\subsection{The components ${\si}_{\rho \rho}$ and 
${\si}_{\phi \phi}$ }

In the previous section we found ${\cF}(\rho)$ which respects
the inhomogeneous gauge equation (3.36) with the source
term ${\cR}(\rho)$ taken in the approximate form (3.44). The 
function ${\cF}(\rho)$ implies the difference 
$f(\rho)-f_{{\rm bg}}(\rho)$, where $f(\rho)$ is the modified stress 
potential of second order, and $f_{{\rm bg}}(\rho)$ is the 
background stress potential. Practically, ${\cF}(\rho)$ is given 
by the set of integral representations (4.13), (4.14), and (4.18). 
However, we shall not attempt to elaborate a single formula which 
would express ${\cF}(\rho)$ through ${\cR}(\rho)$ 
more explicitly. The most important for us asymptotical properties 
of the stress field of the modified screw dislocation can be 
obtained just from Eqs. (4.13), (4.18).

Equation (4.18) allows to express the modified stress 
potential $f$ as follows:
$$
f\,=\,f_{{\rm bg}}\,+\,C\,{\w Y}_0 ({\cN}\rho)\,+\,I_{{\cF}}\,,
\eqno(5.1.1)
$$
$$
f_{{\rm bg}}\,=\,-k\,\frac{p_1}{8\kappa^2}\,\log^2\rho
\,+\,d_1\,\rho^2\,+\,d_2\,\log \rho\,,
\eqno(5.1.2)
$$
where the classical second order stress potential 
$f_{{\rm bg}}$ (5.1.2) is written in the form suggested by 
\cite{pfl}. Free parameters $d_1$ and $d_2$ in (5.1.2) are  
determined in \cite{pfl} from the requirement that the second 
order stress $\stackrel{(2)}{\si}_{\rho \rho}$ vanishes at the 
boundaries of a hollow cylinder $\rho=\rho_c$ and $\rho=\rho_e 
> \rho_c$. In what follows, it will be seen that the cut-off 
at $\rho=\rho_c$ disappears in the gauge model proposed.

The second order stress tensor of the modified screw
dislocations is given by the following relations:
$$
\stackrel{(2)}{\si}_{\rho \rho}\,=\,
-\,\frac{1}{\rho}\,\frac{d}{d \rho}\,f\,,\qquad
\stackrel{(2)}{\si}_{\phi \phi}\,=\,
-\,\frac{d^2}{d \rho^2}\,f\,,\qquad
\stackrel{(2)}{\si}_{z z}\,=\,p\,,
\eqno(5.2.1)
$$
where $f$ is given by (5.1) and $p$ is given by (3.35).
Other components of $\stackrel{(2)}{{\bsi}}$ are zero.
It is most important for us to consider the limiting
behaviour of the solution (5.1), (5.2) at ${\cN}\rho\gg 1$
and ${\cN}\rho\ll 1$. Let us begin with the case

\vskip 0.5cm
\leftline{A) $\quad {\cN}\rho\ll 1$}
\vskip 0.5cm

Since our attention is attracted now to 
$\stackrel{(2)}{\si}_{\rho \rho}$ and 
$\stackrel{(2)}{\si}_{\phi \phi}$, let us write, using
(4.19) and (5.1), their expansions as follows:
$$
\begin{array}{rcl}
\stackrel{(2)}{\si}_{\rho \rho}\,=&-&
\displaystyle{
2 \Bigl(r_2\,-\,k\frac{p_1}{8\kappa^2}\Bigr)\,\frac{\log\rho}{\rho^2}
-(r_1+d_2)\,\frac{1}{\rho^2}}\,-\,2 r_3\log^3\rho\,-\,
(3 r_3+2 r_4) \log^2\rho \\[0.5cm]
&-&2 (r_4+r_5)\log\rho\,-\,(r_5+2r_6+2d_1)\,-\,
4 r_7 \rho^2\log^3\rho\,,\\ [0.5cm]
\stackrel{(2)}{\si}_{\phi \phi}\,=& &
\displaystyle{
2 \Bigl(r_2\,-\,k\frac{p_1}{8\kappa^2}\Bigr)\,\frac{\log\rho}{\rho^2}
-\Bigl(2(r_2\,-\,k\frac{p_1}{8\kappa^2})\,-\,r_1\,-\,d_2\Bigr)\,
\frac{1}{\rho^2}} \\ [0.5cm]
&-& 2 r_3\log^3\rho\,-\,(9 r_3\,+\,2 r_4) \log^2\rho\,-\,
2 (3 r_3\,+\,3 r_4\,+\,r_5)\log\rho \\[0.5cm]
&-&(3 r_5\,+\,2 r_4\,+\,2 r_6\,+\,2 d_1)\,-\,12 r_7 
\rho^2\log^3\rho\,.
\end{array}
\eqno(5.2.2)
$$
It was noticed in Subsection 4.1 that for our choice of the density 
profile, the series expansion ${\cR}$ (4.9) at small distances is 
missing the term corresponding to the coefficient $p_2$. In its turn, 
the vanishing of $p_2$ just implies absence of the terms depending on 
$r_3$ and $r_7$ in (4.19) and so in (5.2.2). Besides, the constant 
term $r_0$ is irrelevant for the stress components 
$\stackrel{(2)}{\si}_{\rho \rho}$, 
$\stackrel{(2)}{\si}_{\phi \phi}$ (5.2.1). The contribution
corresponding to $r_2$ is compensated exactly by the first 
term in $f_{{\rm bg}}$ (5.1.2) (and thus the contribution
$\propto\rho^{-2}\log\rho$ disappears in (5.2.2)).

Therefore, our attention should be paid only to the coefficients 
containing $r_1$, $r_4$, $r_5$, $r_6$ in the final expansions 
(5.2.2). First of all, it is interesting that there exists an 
opportunity to make $r_4$ equal to zero: $r_4=0$. Indeed, using $r_4$ 
in the form (4.20) (where we put $p_2=0$ and use $p_1$, $p_3$ in 
the form given by (4.10)), we can re-express equation $r_4=0$ as 
follows:
$$
k\,\frac{X^2}8\,\left(
\frac{2{\w c}}{z^2_c}\,-\,c\,-\,\frac{{\w c}}4\Bigl(
1\,-\,\frac{{\cN}^2}{\kappa^2}\Bigr)\right)\,=\,0\,,
\eqno(5.3)
$$
or, after the use of ${\w c}$ in the form (3.42), 
$$
c\,=\,\left(\frac{1-2 \mu^2 C_7}{1+\nu}
\,-\,4 c\right)\,\left(\frac{\nu}{2(1-\nu)}\,+\,
\frac{2}{z^2_c}\right)\,.
\eqno(5.4)
$$
Let us rewrite (5.4) using $c$ (3.39) as follows:
$$
\frac{2a{\cZ}\,+\,1}{{\cZ}\,+\,1}\,=\,\eta\,\equiv\,
\frac{4\mu^2 (1\,+\,\nu) C_3}{1\,-\,2\mu^2 C_7}\,,
\qquad {\cZ}\,\equiv\,\frac{1+\nu}{1-\nu}\times
\frac{z^2_c}{8}\,,
\eqno(5.5)
$$
where $a\equiv \nu/(1+\nu)$. Left-hand side
of (5.5) is positive (it is known that $0 <\nu\le 1/2$ for 
isotropic materials), and thus $\eta$ in its R.H.S. is
also positive. Therefore, the following two requirements 
appear: 
$$
a)\quad C_3\,>\,0\,,\quad C_7\,<\,1/2\mu^2\,,
$$
$$
b)\quad C_3\,<\,0\,,\quad C_7\,>\,1/2\mu^2\,.
$$
Provided these requirements are fulfilled, Eq. (5.5) can be solved 
for ${\cZ}$:
$$
{\cZ}\,=\,\frac{1\,-\,\eta}{\eta\,-\, 2 a}\,.
\eqno(5.6)
$$
In its turn, Eq. (5.6) results in the following 
restrictions on the parameters $C_3$, $C_7$:
$$
\begin{array}{l}
a)\quad \displaystyle{
C_7\,+\,2 (1+\nu)\,C_3\,<
\frac{1}{2\mu^2}\,<\,C_7\,+\,\frac{(1+\nu)^2}{\nu}\,C_3}\,,\\
b)\quad \displaystyle{
C_7\,+\,\frac{(1+\nu)^2}{\nu}\,<
\frac{1}{2\mu^2}\,<\,C_7\,+\,2 (1+\nu)\,C_3}\,.
\end{array}
\eqno(5.7)
$$
(Notice that $(1+\nu)/\nu\ge 3$ at $0 <\nu\le 1/2$.)
For instance, at $\nu= 1/3$, Eqs. (5.7) take the form:
$$
\begin{array}{l}
a)\quad \displaystyle{
C_7\,+\,\frac83\,C_3\,<
\frac{1}{2\mu^2}\,<\,C_7\,+\,\frac{16}3\,C_3}\,,\\[0.5cm]
b)\quad \displaystyle{
C_7\,+\,\frac{16}3\,C_3\,<
\frac{1}{2\mu^2}\,<\,C_7\,+\,\frac83\,C_3}\,,
\end{array}
\eqno(5.8)
$$
and
$$
\frac{z_c}{2}\,=\,\left(\frac{1\,-\,\eta}{\eta\,-
\,1/2}\right)^{1/2}
\eqno(5.9)
$$
in both cases. 
Provided Eqs. (5.8) are fulfilled, $\eta$ respects $1/2 <\eta < 1$, 
and the parameter $z_c$ (5.9) can formally acquire
any real positive value.

Therefore, the requirement $r_4=0$ turns out to be highly
interesting since it gives us a formal expression for the 
radius of the domain of localization of the gauge dislocation's density 
profile (just under our approximation (3.29)).

Now let us focus at the coefficients $r_1$, $r_5$, $r_6$. 
Here it is appropriate to impose the following constraints
(notice that ${\h J}_4={\h Y}_4$ since $p_2=0$):
$$
\begin{array}{l}
r_1\,+\,d_2\,=\,C\,+\,{\w {\cal I}_J}\,+\,d_2\,=\,0\,,
\\[0.5cm]
\displaystyle{
r_5\,=\,{\h J}_6\,-\,{\h Y}_6\,+\,\log\Bigl({\cN}^2\Bigr)
\bigl({\h J}_5\,-\,{\h Y}_5 \bigr)
\,-\,\frac{{\cN}^2}4\,C\,=\,0}\,,\\[0.5cm]
\displaystyle{
r_6\,+\,d_1\,=\,{\h J}_7\,-\,{\h Y}_7\,+\,\log{\cN}\,
\bigl({\h J}_6\,-\,{\h Y}_6 \bigr)\,+\,\log^2{\cN}\,
\bigl({\h J}_5\,-\,{\h Y}_5 \bigr)}\\[0.5cm]
\displaystyle{
\qquad\qquad +\,C\,\frac{{\cN}^2}4
\left(1\,-\,\log\Bigl(\frac{\ga}2{\cN}\Bigr)\right)
\,+\,d_1\,=\,0 }\,.
\end{array}
\eqno(5.10)
$$
Equations (5.10) simply express the fact that the terms 
proportional to $\log \rho$, $\rho^2 \log \rho$, and 
$\rho^2$ are absent in $f$ (5.1) at $\rho\ll 1$ (see expansion
(4.19)). Therefore, under our conventions all the terms in 
(5.2.2) vanish. However, possible contributions
$\propto\rho^2\log^2\rho$ should be expected provided $f$
is expanded further.

\vskip 0.5cm
\leftline{B) $\quad {\cN}\rho\gg 1$}
\vskip 0.5cm

In this limit 
\footnote{${\cN}$ is large but $\rho$ is not necessarily large;
validity of the replacement (3.29) suggests that $\kappa\sim {\cN}$
is large},
the `integral' contribution $I_{{\cF}}$ in $f$ (5.1) 
is exponentially small. Therefore, the asymptotics of $f$ at 
some $\rho_e$, ${\cN}\rho_e\gg 1$, is as follows:
$$
\displaystyle{
f\,\simeq\,f_{{\rm bg}} (\rho_e)\,+\,
C\,\biggl(\frac{\pi}{2 {\cN}\rho_e}\biggr)^{1/2}\,
\!\!\!\!\sin\Bigl({\cN}\rho_e\,-\,\frac\pi{4}\Bigr)}\,.
\eqno(5.11)
$$
Using (5.11), we obtain the boundary condition
$\stackrel{(2)}{\si}_{\rho \rho}|_{\rho=\rho_e}=0$, i.e., a free 
surface boundary condition, in the form:
$$
\Bigl.\stackrel{(2)}{\si}_{\rho \rho}\Bigr|_{\rho=\rho_e}\,=\,
\displaystyle{
2 k\,\frac{p_1}{8\kappa^2}\,
\frac{\log \rho_e}{\rho_e^2}\,-\,\frac{d_2}{\rho^2_e}\,-\,
2 d_1\,-\,C\,\frac{{\w Y}_1 ({\cN}\rho_e)}{\rho_e}\,=\,0}\,,
\eqno(5.12.1)
$$
where 
$$
{\w Y}_1 ({\cN}\rho)\equiv 
\displaystyle{
\frac{d {\w Y}_0 ({\cN}\rho)}{d\rho}\,=\,-\,{\cN}\,
\frac{\pi}2\,Y_1({\cN}\rho)}\,.
$$
Besides, under the condition ${\cN}\rho\gg 1$ we obtain:
$$
\stackrel{(2)}{\si}_{\rho \rho}\,+\,
\stackrel{(2)}{\si}_{\phi \phi}\,=\,
\displaystyle{
2 k\,\frac{p_1}{8\kappa^2}\,\frac{1}{\rho^2}\,-\,4 d_1\,+\,
C\,{\cN}^2\,{\w Y}_0 ({\cN}\rho)\,,
}
\eqno(5.12.2)
$$
where the representation for $\stackrel{(2)}{\si}_{\rho \rho}$
is seen from (5.12.1) (with $\rho$ instead of $\rho_e$). When 
(5.12.1) is fulfilled, i.e., at $\rho=\rho_e$, the boundary value of 
$\stackrel{(2)}{\si}_{\phi \phi}$ is given by R.H.S. of (5.12.2) 
provided $\rho$ is replaced by $\rho_e$. It is seen that the 
boundary value of $\stackrel{(2)}{\si}_{\phi \phi}$ tends to 
$-4d_1$ at $\rho_e\to\infty$.

Let us note furthermore that the second equation in (5.10), i.e., 
the condition $r_5=0$, can also be fulfilled separately since 
the difference
${\h J}_6\,-\,{\h Y}_6$ depends on the coefficient $q_4$ (4.16), 
which, in turn, contains another free parameter (denoted above as 
$const$) to be adjusted. Provided $r_5=0$ is fulfilled by a choice 
of $const$, the following three equations, the first and the third 
Eqs. (5.10), and Eq. (5.12.1), can be written together as a 
single $3\times 3$ matrix equation as follows:
$$
\left(
\begin{array}{ccc}
0 & 1& 1\\
1 & 0& {\cM}_1\\
2\rho^2_e & 1 & {\cM}_2
\end{array}\right)
\left(
\begin{array}{c}
d_1 \\d_2\\  C
\end{array}\right)\,=\,
\left(\begin{array}{c}
l_1 \\l_2\\ l_3
\end{array}\right)\,,
\eqno(5.13)
$$
where
$$
\begin{array}{l}
l_1\,\equiv\,-\,{\w {\cal I}_J}\,,\\[0.5cm]
l_2\,\equiv\,-\,\bigl({\h J}_7\,-\,{\h Y}_7\bigr)
\,-\,\log{\cN}\,\bigl({\h J}_6\,-\,{\h Y}_6 \bigr)\,-\,
\log^2{\cN}\,\bigl({\h J}_5\,-\,{\h Y}_5 \bigr)\,,\\[0.5cm]
\displaystyle{
l_3\,\equiv\,k\,\frac{p_1}{4\kappa^2}\,\log \rho_e}\,,\\[0.5cm]
\displaystyle{
{\cM}_1\,\equiv\,\frac{{\cN}^2}4
\left(1\,-\,\log\Bigl(\frac{\ga}2{\cN}\Bigr)\right)}\,,\\[0.5cm]
\displaystyle{
{\cM}_2\,\equiv\,-\,\frac{\pi}2\bigl({\cN}\rho_e\bigr)\,
Y_1 ({\cN}\rho_e)\,\simeq\,\left(\frac{\pi}2 {\cN}\rho_e\right)^{1/2}
\cos\Bigl({\cN}\rho_e\,-\,\frac\pi{4}\Bigr)}\,.
\end{array}
$$

The determinant $(\equiv {\cD})$ of the matrix in L.H.S. of (5.13),
$$
\begin{array}{rcl}
{\cD}&=&2{\cM}_1\,\rho^2_e\,-\,{\cM}_2\,+\,1 \\ [0.5cm]
&=& \displaystyle{
\left(1\,-\,\log\Bigl(\frac{\ga}2{\cN}\Bigr)\right)
\frac{\bigl({\cN}\rho_e\bigr)^2}2\,+\,\frac\pi{2}\,{\cN}\rho_e
\,Y_1 ({\cN} \rho_e)}\,+\,1\,,
\end{array}
\eqno(5.14)
$$
is non-zero. Therefore, Eqs. (5.13) can be solved for 
$d_1$, $d_2$, $C$, and by the Cramer's rule we obtain:
$$
\begin{array}{l}
d_1\,=\,{\cD}^{{\1}}\,
\Bigl(\bigl(1\,-\,{\cM}_2\bigr)l_2\,+\,
{\cM}_1\bigl(l_3\,-\,l_1\bigr)\Bigr)\,,\\[0.5cm]
d_2\,=\,{\cD}^{{\1}}\,
\Bigl(\bigl(2 \rho_e^2 {\cM}_1-{\cM}_2\bigr) l_1\,-\,
2 \rho_e^2\,l_2\,+\,l_3\Bigr)\,,\\[0.5cm]
C\,=\,{\cD}^{{\1}}\,\bigl(l_1\,+\,2\,\rho^2_e\,l_2\,-\,l_3\bigr)\,.
\end{array}
\eqno(5.15)
$$
It should be noticed that Eqs. (5.14), (5.15) can be simplified 
at ${\cN} \rho_e\gg 1$.

Therefore, all free constants at our disposal, i.e., $d_1$, $d_2$ 
(see (5.1)), $C$ (see (4.18)), $const$ (see (4.13.2)), 
$z_c=\kappa\rho_c$ (see (3.29)), ${\cC}^\prime$ and ${\cC}^{\prime 
\prime}$ (see (3.30)) are fixed. So far, only $Q^\prime$ (3.23) 
remains to be chosen. It is remarquable that the approach developed 
allows to determine the core radius 
$\rho_c=\displaystyle{\frac{z_c}{\kappa}}$ as a function of two elastic 
moduli of second order (say, $\mu$ and $\nu$) and of two third order 
elastic moduli $C_3$ and $C_7$ (under our assumption about the 
simplified density profile). For the given choice of the parameters, 
all the contributions in $f$ (5.1) at $\rho\ll 1$ up to those 
$\propto\rho^4\log^3\rho$ (apart from the irrelevant constant 
$r_0$) are zero. This means that leading non-vanishing contribution 
to $\stackrel{(2)}{\si}_{\rho \rho}$, $\stackrel{(2)}{\si}_{\phi \phi}$ 
should be expected as $\propto\rho^2\log^2\rho$ at $\rho\to 0$. The
corresponding term can be deduced explicitly after a straightforward 
calculation: the contribution $\propto\rho^4\log^2\rho$ 
in $\stackrel{(2)}{f}$ must be accounted for in that case. Thus,
$\stackrel{(2)}{\si}_{\rho \rho}$, $\stackrel{(2)}{\si}_{\phi \phi}$ 
tend to zero in our model at $\rho\to 0$. Besides, 
$\stackrel{(2)}{\si}_{\rho \rho}$ is zero at the external boundary 
$\rho=\rho_e$.

\subsection{The component $\si_{z z}$}

Eventually, let us consider the stress component
$\si_{z z}=p$. For the background contribution,
we obtain from (3.24):
$$
\displaystyle{
\bigl(\si_{{\rm bg}}\bigr)_{z z}\,=\,-
\frac{a}{1-a}\,\Dl f_{{\rm bg}}\,-\,\frac{2\mu}{1-a}
\bigl(\Psi^{{\rm bg}}_{z z}\,+\,Q^\prime\bigr)\,.
}
\eqno(5.16)
$$
In the classical approach we consider 
$\bigl(\si_{{\rm bg}}\bigr)_{z z}$ (5.16)
at $\rho_c\le\rho\le\rho_e$, and we determine
$Q^\prime$ from the requirement 
$$
\int\limits_0^{2\pi}\int\limits_{\rho_c}^{\rho_e}
\si_{z z}\,\rho\,d\rho\,d\varphi\,=\,0\,,
\eqno(5.17)
$$
which is to express that a ``mean'' value of $\si_{z z}$ 
(i.e., $\si_{z z}$ averaged over cross-section of the bulk)
is zero. So, in order to determine $Q^\prime$, we take 
$f_{{\rm bg}}$ in the form given by (5.1.2), and we also
take into account Eqs. (3.21), (3.24) where 
$\stackrel{(1)}{\phi}_{{\rm bg}}\equiv (-b/2\pi) \log\rho$ 
is used for evaluation of $\Psi^{{\rm bg}}_{z z}$.
Then, using (5.16) and (5.17) we find an expression
for $Q^\prime$. Substituting this $Q^\prime$ into (5.16) 
we obtain:
$$
\displaystyle{
\bigl(\si_{{\rm bg}}\bigr)_{z z}\,=\,
\left[\biggl(\frac{b}{2\pi}\biggr)^2
\frac{2\mu^3 C_3}{1-a}\,+\,k\,\frac{p_1}{4\kappa^2}\,
\frac{a}{1-a}\right]\,\left[\frac1{\rho^2}\,-\,
\frac2{\rho_e^2-\rho_c^2}\,\log\frac{\rho_e}{\rho_c}\right]\,,
}
\eqno(5.18)
$$
where $\nu=a (1-a)^{\1}$ and $(1-a)^{\1}=1+\nu$.

Using Eqs. (3.8) and (3.9), one can re-express the coefficient
$\displaystyle{k\,\frac{p_1}{4\kappa^2}}$
(where $p_1$ is given by (4.10) with ${\w c}$ in it given by 
(3.42), and $X^2$ is introduced in (4.2)) as follows:
$$
\displaystyle{
k\,\frac{p_1}{4\kappa^2}\,=\,-\biggl(\frac{b}{2\pi}\biggr)^2\,
\frac{\mu}{2(1-\nu)}\,\left[
\frac{n^\prime}{4\mu}\,-\,4 \mu^2 (1+\nu)\,C_3\right]\,,
}
\eqno(5.19.1)
$$
$$
\displaystyle{
=\,-\biggl(\frac{b}{2\pi}\biggr)^2\,\left(
\mu\,+\,\frac{1-2\nu}{1-\nu}\,
\frac{2 m^\prime\,+\,n^\prime}8 \right)\,.
}
\eqno(5.19.2)
$$
Furthemore, let us express $C_3$ by means of (5.19.1) and then 
substitute it into (5.18). Now $\bigl(\si_{{\rm bg}}\bigr)_{z z}$
acquires the form:
$$
\displaystyle{
\bigl(\si_{{\rm bg}}\bigr)_{z z}\,=\,
\left(k\,\frac{p_1}{4\kappa^2}\,+\,
\biggl(\frac{b}{2\pi}\biggr)^2 \frac{n^\prime}8\right)
\,\left(\frac1{\rho^2}\,-\,
\frac2{\rho_e^2-\rho_c^2}\,\log\frac{\rho_e}{\rho_c}\right)
}\,.
\eqno(5.20.1)
$$
Eventually, we use (5.19.2) to re-express 
$\displaystyle{k\,\frac{p_1}{4\kappa^2}}$ in (5.20), and
the stress component $\bigl(\si_{{\rm bg}}\bigr)_{z z}$
appears in the form suggested by \cite{pfl}:
$$
\displaystyle{
\bigl(\si_{{\rm bg}}\bigr)_{z z}\,=\,-\,\biggl(\frac{b}{2\pi}\biggr)^2
\left[\mu\,+\,\frac1{4(1-\nu)}\Bigl(m^\prime (1-2\nu)\,-\,
n^\prime \frac{\nu}2\Bigr)\right]
\,\left(\frac1{\rho^2}\,-\,
\frac2{\rho_e^2-\rho_c^2}\,\log\frac{\rho_e}{\rho_c}\right)
}\,.
\eqno(5.20.2)
$$

It is also important to notice the fact that the integral
$$
\int\limits_{\rho_c}^{\rho_e}\Dl f_{{\rm bg}}\,\rho\,d\rho
\eqno(5.21) 
$$
is zero, since the choice of $d_1$, $d_2$ in $f_{{\rm bg}}$ 
(5.2) ensures vanishing of $d f_{{\rm bg}}/d\rho$ at
$\rho=\rho_e, \rho_c$ (the boundary condition for
$\stackrel{(2)}{\si}_{\rho \rho}$ at $\rho=\rho_e, 
\rho_c$). Therefore, $\bigl(\si_{{\rm bg}}\bigr)_{z z}$
can also be written in the following form:
$$
\bigl(\si_{{\rm bg}}\bigr)_{z z}\,=\,
-\nu\,\Dl f_{{\rm bg}}\,-\,2\mu (1+\nu)
\left[\Psi^{{\rm bg}}_{z z}\,-\,\frac2{\rho_e^2-\rho_c^2}
\,\int\limits_{\rho_c}^{\rho_e}\Psi^{{\rm bg}}_{z z}\,
\rho\,d\rho\right]\,.
\eqno(5.22)
$$
Equation (5.22) is completely equivalent to (5.18).
In view of (5.22), it is more clear, how Eq. (5.17) 
is fulfilled.

Now let us turn to $\si_{z z}$ of the modified defect
given by (3.35):
$$
\displaystyle{
\si_{z z}\,=\,-\nu\,\Dl f\,-\,2\mu\,(1+\nu) 
(\Psi_{3 3}\,+\,{\w g})\,+\,\kappa^2\,(1+\nu)\,(f-f_{{\rm bg}})
}\,.
\eqno(5.23)
$$
Let us consider the asymptotical properties of $\si_{z z}$.
Using the expressions for $\stackrel{(2)}{\si}_{\rho \rho}$,
$\stackrel{(2)}{\si}_{\phi \phi}$ (5.2.2), the expansion (4.19),
and $\stackrel{(1)}{\phi}$ to express $\Psi_{3 3}$ (as well as
the nullification conditions (5.10)), we obtain that $\si_{z z}$
behaves as $\propto (r_1\,\log\rho+r_2\,\log^2\rho)$ plus 
the contribution tending to zero at $\rho\to 0$. 
At ${\cN}\rho\gg 1$ we obtain:
$$
\si_{z z}\,\simeq\,-4 \nu d_1\,+\,
\displaystyle{
\frac{1+\nu}{1-\nu}\,
\left[ C {\kappa}^2 {\w Y}_0\,+\,\biggl(\frac{b}{2\pi}\biggr)^2
\,\mu^3\,\left(2C_3\,+\,\frac{\nu}{1+\nu}
\Bigl(C_7\,-\,\frac{1}{2\mu^2}\Bigr)\right)\,\frac{1}{\rho^2}
\right]\,.
}
\eqno(5.24)
$$
With the help of (5.12.2) and (5.24) we obtain the
trace of $\stackrel{(2)}{{\bsi}}$ (and thus the total trace, since
$\stackrel{(1)}{{\bsi}}$ for the screw dislocation is traceless)
at ${\cN}\rho\gg 1$:
$$
\begin{array}{rcl}
\tr (\stackrel{(2)}{{\bsi}})&\equiv&I_1(\stackrel{(2)}{{\bsi}})
\,=\,-4 (1\,+\,\nu) d_1 \\ [0.5cm]
&+&
\displaystyle{
\frac{1+\nu}{1-\nu}\,
\left[ 2 C {\kappa}^2 {\w Y}_0\,+\,\biggl(\frac{b}{2\pi}\biggr)^2
\,\mu^3\,\left(4C_3\,+\,C_7\,-\,\frac{1}{2\mu^2}\right)\,
\frac{1}{\rho^2}\right]\,.}
\end{array}
\eqno(5.25)
$$
Besides, $C$ and $d_1$ given by (5.15) must be substituted into
(5.24) and (5.25). Equations (3.6) and (3.21) demonstrate us how 
an analogous estimation for the trace of $\stackrel{(2)}{{\bf e}}$
can be deduced. Indeed,
$$
\begin{array}{rcl}
I_1 (\stackrel{(2)}{{\bf e}})&=&(3 C_1\,+\,C_4)\,
I_1(\stackrel{(2)}{{\bsi}})\,+\,I_1 (\stackrel{(1)}{{\bPsi}})\\[0.5cm]
&=&(3 C_1\,+\,C_4)\,I_1(\stackrel{(2)}{{\bsi}})\,-\,
\mu^2\,(3 C_3\,+\,C_7)\,\Bigl(\cd_\rho\stackrel{(1)}{\phi}\Bigr)^2\,.
\end{array}
\eqno(5.26)
$$
However, at $\kappa\rho\gg 1$, we estimate 
$(\cd_\rho\stackrel{(1)}{\phi})^2\simeq b^2/(2\pi\rho)^2$,
and the trace of $\stackrel{(2)}{{\bf e}}$ can be deduced by
means of (5.25) and (5.26).

Eventually, using
$$
{\w g}\,=\,Q^\prime\,+\,\frac{b^2}{8\pi^2\rho^2_c}\,
\left(\frac{\rho^2}{\rho^2_c}\,-\,1\right)\,
h_{[0,\,\rho_c]}(\rho)\,,
$$
and the fact that $\rho\,\Dl \stackrel{(2)}{f}$ 
integrated over $\rho$ from $0$ to $\rho_e$ vanishes,
we determine $Q^\prime$. Substituting $Q^\prime$
into (5.23), we find:
$$
\begin{array}{rcl}
\si_{z z}\,= &-&
\displaystyle{
\nu\,\Dl f\,-\,2\mu\,(1+\nu) 
\left[\Psi_{3 3}\,-\,\frac2{\rho_e^2}
\,\int\limits_0^{\rho_e}\Psi_{3 3}\,\rho\,d \rho\right]} \\[0.5cm]
&-&
\displaystyle{
2\mu\,(1+\nu)\left[\frac{b^2}{16 \pi^2 \rho^2_e}\,+\,
\frac{b^2}{8\pi^2\rho^2_c}\,
\left(\frac{\rho^2}{\rho^2_c}\,-\,1\right)\,
h_{[0,\,\rho_c]}(\rho)\right]
} \\[0.5cm]
&+&
\displaystyle{
\kappa^2\,(1+\nu)
\,\left[ f-f_{{\rm bg}}\,-\,\frac2{\rho_e^2}
\,\int\limits_0^{\rho_e}\bigl(f-f_{{\rm bg}}\bigr)
\,\rho\,d \rho\right]
}\,.
\end{array}
\eqno(5.27)
$$
We shall not elaborate this expression further. It is
enough to notice that the integrals in (5.27) are convergent at
the lower bands, and thus $\si_{z z}$ (5.27) averaged over
cross-section of the bulk is zero.

Before to conclude, let us briefly note another possibility
which concerns the choice of the parameters. Namely, the
requirement $r_6+d_1=0$ can be left aside. In this case,
the stresses $\stackrel{(2)}{\si}_{\rho \rho}$,
$\stackrel{(2)}{\si}_{\phi \phi}$ (5.2.2) will demonstrate a tending,
at $\rho\to 0$, to the constant values $\sim (r_6+d_1)$ (see (5.2.2)). 
However, in this case it can be assumed that $C=0$. Then, instead of 
(5.13), we shall get just two equations to determine $d_1$ and $d_2$. 
First, we obtain $d_2=l_1\equiv -{\w {\cal I}_J} = -r_1$. Then, Eq.
(5.12.1) takes the form:
$$
\displaystyle{\frac{l_3}{\rho^2_e}\,-\,\frac{d_2}{\rho^2_e}}
\,-\,2 d_1\,=\,0\,,
$$
and it gives us
$$
\displaystyle{
d_1\,=\,\frac{l_3\,-\,d_2}{2\rho^2_e}}\,=\,
\displaystyle{
k\,\frac{p_1}{8\kappa^2}\frac{\log\rho_e}{\rho_e^2}
\,+\,\frac{{\w {\cal I}_J}}{2\rho_e^2}\,.
}
$$
In this case, the asymptotical behaviour of 
$\stackrel{(2)}{\si}_{\rho \rho}$, $\stackrel{(2)}{\si}_{\phi \phi}$
is missing the unconventional contribution due to ${\w Y}_0$.
Let us stress again that the parameters $d_1$ and $d_2$ are
still different in comparison with the analogous conventional results.

\section{ Discussion}

A model of non-singular screw dislocation lying along an infinitely 
long cylindric body is investigated in the present paper in the 
framework of three-dimensional $\T$-gauge approach \cite{mal}. 
The gauge part of the total Lagrangian is chosen in 
the Hilbert--Einstein form, while the elastic contribution to it
corresponds to the energy of elastically isotropic continuum given 
by the terms of second and third orders in the strain components. 
In other words, a second order elasticity approach is adopted in 
the present paper.

As it was noticed in \cite{mal}, second order consideration
in the framework of the model \cite{mal} would merit attention as an 
attempt to clarify perspectives of such rather non-traditional approach 
to defects in solids as the gauge Lagrangian approach (based, for 
instance, on the groups either $T(3)$ or $\ISO\equiv T(3)\ppu SO(3)$). 
Elaboration of related technical details could clarify the gauge 
strategy itself concerning a choice of Lagrangian's constituents, of 
dimensionality of the specific problems, of resolving ansatz, etc. 
On the other hand, it is also intriguing to use such a widely
acknowledged and fruitful method as the stress function approach 
\cite{ks}, \cite{pfl} within an unconventional non-linear gauge 
framework. Although the available gauge solutions of the first order 
\cite{ed3}, \cite{mal}, \cite{laz1}, \cite{laz2}, \cite{laz3}, 
\cite{laz4} seem to be promising, second order consideration could 
open new aspects of the problem of the gauge description of dislocations. 

Let us remind that the linear approach developed in \cite{mal} 
leads to the modifed defects which are characterized by the 
fact that singularities of the ordinary straight dislocations are 
smoothed out. After the classical attempts \cite{ks}, 
\cite{pfl}, \cite{seegm}, \cite{wseeg}, higher corrections 
to the law $1/\rho$ of linear elasticity are known. However, a 
cut-off near the dislocation axis inevitably occures 
\cite{pfl}, \cite{seegm}. As it is shown above, the gauge approach 
\cite{mal} allows to extend the description of static screw 
dislocation to the whole cylindric body containing the defect. 
A use of an approximated density-profile comes to play, and an 
expression for the radius of the domain of localization of the
defect's density profile by means of the second and third 
order elastic moduli appears in the picture proposed in the present 
paper.

Second order consideration developed above allows to avoid a 
stress-free boundary condition at an inner radius corresponding to the 
core radius $\rho=\rho_c$. Besides, it allows to fix the radius 
$\rho=\rho_c$ as a function of second and third order elastic moduli. 
It removes a cut-off which occures in \cite{pfl}, and the stress 
components $\si_{\rho \rho}$, $\si_{\phi \phi}$ turn out to be 
continuable towards the tube's axis. As in the classical approach, 
it is necessry to subject $\stackrel{(2)}{\bsi}$ to a free-surface 
boundary condition at the outer radius $\rho=\rho_e$. Thus we obtain 
the solution describing a finite cylinder with nonsingular screw 
dislocation along its axis. Sufficiently far from the core, the 
analytical form of the gauge stress potential of second order found 
above is rather close to the conventional one \cite{pfl}. 

Two possibilities are pointed out for the choice of the parameters
in the solution found. These possibilities enable
$\stackrel{(2)}{\si}_{\rho \rho}$, $\stackrel{(2)}{\si}_{\phi \phi}$
to tend either to constant values or to zero. In the first case,
the analytical form of $\stackrel{(2)}{\si}_{\rho \rho}$,
$\stackrel{(2)}{\si}_{\phi \phi}$ in the region $\rho_c\le\rho\le\rho_e$
is the same as in \cite{pfl}, but the coefficients are nevertheless 
different. In the second case, an unconventional contribution
is present. However, $\stackrel{(2)}{\si}_{z z}$ is logarithmically
divergent (the classical divergency is $\sim\rho^{-2}$), at
$\rho\to 0$, i.e., in the region where the approximated form of the density
profile is most inadequate. In the last case, a weak three-dimensionality
may be of help. Some estimations which involve the crystallographic 
parameters are desirable to make contact with the known interpretations 
of the characteristic length $\kappa^{-1}$ in terms of interatomic spacing 
\cite{laz2} (translational gauging), \cite{cem} (non-local elasticity).

A gauge approach close to ours has been proposed in the series of 
papers \cite{laz1}, \cite{laz2}, \cite{laz3}, \cite{laz4},  
which is based on the $\T$-gauge Lagrangian written as 
a combination of the terms quadratic in the torsion components. 
As to the elastic Lagrangian, it is written in \cite{laz1}, 
\cite{laz2}, \cite{laz3} without third order terms
(since only the linear problems are studied). However,
it is proposed in \cite{laz4} to use the terms in the 
Lagrangian which are related to the energy potential of 
the rotation gradients. Incorporation of such terms in \cite{laz4} 
allows to improve the solution found in \cite{mal} for that modified 
defect which demonstrates how the singularity inherent to the classical 
edge dislocation is smoothed out. In the far field, the stress components 
found in \cite{laz4} correctly reproduce those of the edge dislocation.
The Hilbert--Einstein Lagrangian is highly suggestive representative
among the gauge Lagrangians of the differential--geometric origin.
It leads to the self-contained pictures for the modified defects which 
avoid the singularities of the convetntional solutions. However,
the contributions of mechanical origin also merit consideration, and
further efforts in this direction are also needed.

\section*{Acknowledgement}

The research described has been supported in part by RFBR,
Project No. 01-01-01045

\newpage
\section*{Appendix A}

Appendices A and B provide some intermediate results which are 
helpful in obtaining the final asymptotical expressions for the 
modified stress potential.

First of all, we directly obtain expansions 
for $t^{-1} K_0(t) {\cR}(t)$ and $t^{-1} I_0(t) {\cR}(t)$ at $t\ll 1$
\cite{mag}:
$$
\begin{array}{rcl}
t^{-1} I_0(t) {\cR}(t) &\simeq& 
\displaystyle{
p_1\,t^{-3}\,+\,p_2\,t^{-1}
\log t\,+\,\left(\frac{p_1}4\,+\,p_3\right) t^{-1}
}          \\[0.5cm]
&+&
\displaystyle{
\left( p_4 \log^2 t\,+\,\Bigl(\frac{p_2}4\,+\,
    p_5\Bigr)\,\log t\,+\,{\h k} \right) t\,+\,\dots
}\,;
\end{array}
\eqno({\rm A}1)
$$
$$
\begin{array}{rcl}
t^{-1} K_0(t) {\cR}(t) &\simeq& 
\displaystyle{
- p_1\,\log\Bigl(\frac{\ga}2\,t\Bigr)\,t^{-3}\,-\,
\left( p_2 \log^2 t\,+\,k_1\,\log t\,+\,k_2 \right) t^{-1}
}          \\[0.5cm]
&-&
\displaystyle{
\left( p_4 \log^3 t\,+\,k_3 \log^2 t\,+\,k_4\,\log t\,+\,
k_5 \right) t\,+\,\dots
}\,.,
\end{array}
\eqno({\rm A}2)
$$
where the coefficients $p_1$, $p_2$, $p_3$, $p_4$, and 
$p_5$ are given by (4.10), and the coefficients $k_1$, $k_2$, 
$k_3$, $k_4$, $k_5$, are expressed by means of $p_1, \dots, p_5$ 
as follows:
$$
\begin{array}{l}
\displaystyle{
k_1\,=\,\frac{p_1}4\,+\,\log\Bigl(\frac\ga{2}\Bigr)\,p_2\,+\,p_3
}\,,\\[0.5cm]
\displaystyle{
k_2\,=\,-\frac{p_1}4\,+\,\log\Bigl(\frac\ga{2}\Bigr)\,
\left( \frac{p_1}4\,+\,p_3\right)
}\,,\\[0.5cm]
\displaystyle{
k_3\,=\,\frac{p_2}4\,+\,\log\Bigl(\frac\ga{2}\Bigr)\,p_4\,+\,p_5
}\,,\\[0.5cm]
\displaystyle{
k_4\,=\,{\h k}\,-\,\Bigl(1\,-\,\log\frac\ga{2}\Bigr)
\,\frac{p_2}4\,+\,\log\Bigl(\frac\ga{2}\Bigr)\,p_5
}\,,\\[0.5cm]
\displaystyle{
k_5\,=\,\log\Bigl(\frac\ga{2}\Bigr)\,{\h k}\,-\,\frac3{128}\,p_1\,-\,
\frac{p_3}4
}\,,\qquad
\displaystyle{
{\h k}\,=\,w_c(0)\,+\,\frac{p_1}{64}\,+\,\frac{p_3}{4}\,+\,
p_6
}\,.
\end{array}
\eqno({\rm A}3)
$$

Now let us obtain estimations for the integrals in
(4.13). Using (A1) and (A2) we obtain at $z\ll 1$:
$$
\begin{array}{rcl}
\displaystyle{
B(s)\,-\,\int\limits_s^z I_0(t) {\cR}(t)\,\frac{dt}{t}}
& {\stackrel{\simeq}{_{s\to 0}}}& 
\displaystyle{
{\cI}_0\,+\,{\cI}_1\,z^{-2}\,+\,{\cI}_2\,\log^2 z
\,+\,{\cI}_3\,\log z 
}             \\[0.5cm]
&+& \displaystyle{
\left({\cI}_4\,\log^2 z\,+\,{\cI}_5\,\log z\,+\,
{\cI}_6\right)\,z^2\,+\,\dots\,,
}
\end{array}
\eqno({\rm A}4)
$$
where
$$
\begin{array}{l}
\displaystyle{
{\cI}_0\,=\,const\,-\,\frac{p_1}2\,,\quad
{\cI}_1\,=\,\frac{p_1}2\,,\quad
{\cI}_2\,=\,\frac{-p_2}2\,,\quad
{\cI}_3\,=\,\frac{-p_1}4\,-\,p_3\,,
}\\[0.5cm]
\displaystyle{
{\cI}_4\,=\,\frac{-p_4}2\,,\quad
{\cI}_5\,=\,-\frac{p_2}8\,+\,\frac{p_4\,-\,p_5}2\,,\quad
2\,{\cI}_6\,=\,-{\h k}\,-\,{\cI}_5
}\,;
\end{array}
\eqno({\rm A}5)
$$
and
$$
\begin{array}{rcl}
\displaystyle{
\int\limits_z^\infty K_0(t) {\cR}(t)\,\frac{dt}t}
& \simeq &
{\cK}_0\,+\,{\cK}_1\,z^{-2}\,+\,{\cK}_2\,z^{-2}\,\log z \\[0.5cm]
&+& {\cK}_3\,\log^3 z\,+\,{\cK}_4\,\log^2 z\,+\,
{\cK}_5\,\log z \\[0.5cm]
&+&\left({\cK}_6\,\log^3 z\,+\,{\cK}_7\,\log^2 z\,+\,
{\cK}_8\,\log z\,+\,{\cK}_9
\right)\,z^2\,,
\end{array}
\eqno({\rm A}6)
$$
where
$$
\begin{array}{l}
\displaystyle{
{\cK}_0\,=\,{\cI}_K\,+\,\Bigl(1\,+\,2\,
\log\frac\ga{2}\Bigr)\,\frac{p_1}4\,,\quad
{\cK}_1\,=\,-\,\Bigl(1\,+\,2\,
\log\frac\ga{2}\Bigr)\,\frac{p_1}4\,,
}\\[0.5cm]
\displaystyle{
{\cK}_2\,=\,\frac{-p_1}2\,,\quad {\cK}_3\,=\,\frac{p_2}3\,,\quad
{\cK}_4\,=\,\frac{k_1}2\,,\quad{\cK}_5\,=\,k_2\,,\quad
{\cK}_6\,=\,\frac{p_4}2\,,
}\\ [0.5cm]
\displaystyle{
{\cK}_7\,=\,-\frac34\,p_4\,+\,\frac{k_3}2\,,\quad
{\cK}_8\,=\,\frac{-k_3\,+\,k_4}2\,+\,\frac34\,p_4\,,
}\\[0.5cm]
\displaystyle{
{\cK}_9\,=\,\frac{k_3\,-\,k_4}4\,+\,\frac{k_5}2\,-\,
\frac38\,p_4\,.
}
\end{array}
\eqno({\rm A}7)
$$
Besides, the coefficient ${\cI}_K$ in ${\cK}_0$ is given by the
regularized value of the integral:
$$
\begin{array}{rcl}
{\cI}_K &\equiv& \displaystyle{
\int\limits_1^\infty K_0(t) {\cR}(t)\,\frac{dt}t\,+\,
\int\limits_0^1 \Bigg[ K_0(t) {\cR}(t)\,+\,p_1\,
\log\Bigl(\frac{\ga}2\,t\Bigr)\,t^{-2} \Bigg.}
\\[0.5cm]
&+& \displaystyle{
\Bigg.k_2\,+\,k_1\,\log t\,+\,p_2\,\log^2 t\Bigg]\frac{dt}t\,.
}
\end{array}
\eqno({\rm A}8)
$$           

Using expansions (A4) and (A6), we
obtain the expansions we are interested in:
$$
\begin{array}{rcl}
\displaystyle{
K_0 (z)\,\left[
B(0)\,-\,\int\limits_0^z I_0(t) {\cR}(t)\,\frac{dt}{t}\right]}
& \simeq & 
\displaystyle{
{\h K}_0\,+\,{\h K}_1\,z^{-2}\,+\,{\h K}_2\,z^{-2}\,\log z
}             \\[0.5cm]
&+& {\h K}_3\,\log^3 z\,+\,{\h K}_4\,\log^2 z\,+\,
{\h K}_5\,\log z \\[0.5cm]
&+& \displaystyle{
\left({\h K}_6\,\log^3 z\,+\,{\h K}_7\,\log^2 z\,+\,
{\h K}_8\,\log z\,+\,{\h K}_9\right)\,z^2\,,
}
\end{array}
\eqno({\rm A}9)
$$
where
$$
\begin{array}{l}
\displaystyle{
{\h K}_0\,=\,-\log\Bigl(\frac\ga{2}\Bigr)\,{\cI}_0\,+\,
\Bigl(1\,-\,\log\frac\ga{2}\Bigr)\,\frac{{\cI}_1}4\,,\qquad
{\h K}_1\,=\,-\log\Bigl(\frac\ga{2}\Bigr)\,{\cI}_1\,,\qquad
{\h K}_2\,=\,-{\cI}_1\,,
}\\[0.5cm]
\displaystyle{
{\h K}_3\,=\,-{\cI}_2\,,\qquad
{\h K}_4\,=\,-\log\Bigl(\frac\ga{2}\Bigr)\,{\cI}_2\,-\,{\cI}_3\,,\qquad
{\h K}_5\,=\,-{\cI}_0\,-\,\frac{{\cI}_1}4\,-\,
\log\Bigl(\frac\ga{2}\Bigr)\,{\cI}_3\,,
}\\[0.5cm]
\displaystyle{
{\h K}_6\,=\,-\frac{{\cI}_2}4\,-\,{\cI}_4\,,\qquad
{\h K}_7\,=\,\Bigl(1\,-\,\log\frac\ga{2}\Bigr)\,\frac{{\cI}_2}4
\,-\,\frac{{\cI}_3}4\,-\,\log\Bigl(\frac\ga{2}\Bigr)\,{\cI}_4
\,-\,{\cI}_5\,,
}\\[0.5cm]
\displaystyle{
{\h K}_8\,=\,-\,\frac{{\cI}_0}4\,-\,\frac{{\cI}_1}{64}\,+\,
\Bigl(1\,-\,\log\frac\ga{2}\Bigr)\,\frac{{\cI}_3}4\,-\,
\log\Bigl(\frac\ga{2}\Bigr)\,{\cI}_5\,-\,{\cI}_6\,,
}\\[0.5cm]
\displaystyle{
{\h K}_9\,=\,
\Bigl(1\,-\,\log\frac\ga{2}\Bigr)\,\frac{{\cI}_0}4\,+\,
\Bigl(\frac32\,-\,\log\frac\ga{2}\Bigr)\,\frac{{\cI}_1}{64}
\,-\,\log\Bigl(\frac\ga{2}\Bigr)\,{\cI}_6\,.
}
\end{array}
\eqno({\rm A}10)
$$
Analogously,
$$
\begin{array}{rcl}
\displaystyle{
I_0 (z)\,\int\limits_z^\infty K_0(t) {\cR}(t)\,\frac{dt}{t}}
& \simeq & 
\displaystyle{
{\h I}_0\,+\,{\h I}_1\,z^{-2}\,+\,{\h I}_2\,z^{-2}\,\log z
}             \\[0.5cm]
&+& {\h I}_3\,\log^3 z\,+\,{\h I}_4\,\log^2 z\,+\,
{\h I}_5\,\log z \\[0.5cm]
&+& \displaystyle{
\left({\h I}_6\,\log^3 z\,+\,{\h I}_7\,\log^2 z\,+\,
{\h I}_8\,\log z\,+\,{\h I}_9\right)\,z^2\,,
}
\end{array}
\eqno({\rm A}11)
$$
where
$$
\begin{array}{l}
\displaystyle{
{\h I}_0\,=\,{\cK}_0\,+\,\frac{{\cK}_1}4\,,\qquad
{\h I}_1\,=\,{\cK}_1\,,\qquad
{\h I}_2\,=\,{\cK}_2\,,\qquad{\h I}_3\,=\,{\cK}_3\,,
}\\[0.5cm]
\displaystyle{
{\h I}_4\,=\,{\cK}_4\,,\qquad
{\h I}_5\,=\,\frac{{\cK}_2}4\,+\,{\cK}_5\,,\qquad
{\h I}_6\,=\,\frac{{\cK}_3}4\,+\,{\cK}_6\,,
}\\[0.5cm]
\displaystyle{
{\h I}_7\,=\,\frac{{\cK}_4}4\,+\,{\cK}_7\,,\qquad
{\h I}_8\,=\,\frac{{\cK}_2}{64}\,+\,\frac{{\cK}_5}4
\,+\,{\cK}_8\,,
}\\[0.5cm]
\displaystyle{
{\h I}_9\,=\,\frac{{\cK}_0}{4}\,+\,\frac{{\cK}_1}{64}
\,+\,{\cK}_9\,,
}
\end{array}
\eqno({\rm A}12)
$$

Eventually, we obtain the coefficients characterizing 
the asymptotical behaviour of the combination
$$
\displaystyle{
-\,I_0 (z)\,\int\limits_z^\infty K_0(t) {\cR}(t)\,\frac{dt}{t}
\,+\,K_0 (z)\,\left[
B(0)\,-\,\int\limits_0^z I_0(t) {\cR}(t)\,\frac{dt}{t}\right]}
$$
by summing up the corresponding coefficients given by (A9) and
(A11). We obtain that ${\h K}_2-{\h I}_2=0$
(see (A10) and (A5) for ${\h K}_2$, as well as (A12) and (A7)
for ${\h I}_2$), and the coefficients $q_0, \dots, q_7$ (4.15), 
(4.16) appear as follows:
$$
q_i\,=\,{\h K}_i\,-\,{\h I}_i\,,\qquad {\rm at}\quad i\,=\,0, 1\,;
$$
$$
q_i\,=\,{\h K}_{i+1}\,-\,{\h I}_{i+1}\,,
\qquad {\rm at}\quad i\,=\,2,\,\dots 8\,.
$$
Thus, the solution $G(z)$ given by (4.13), (4.14) is estimated,
and the final answer is given by (4.15), (4.16).

\section*{Appendix B} 

First of all, we obtain expansions for
$t J_0({\cN}t) G(\kappa t)$ and 
$t {\w Y}_0({\cN}t) G(\kappa t)$ at $t\ll 1$:
$$
\begin{array}{rcl}
\displaystyle{
k^{-1}}\!\!\!&t&\!\!\!J_0({\cN}t)\,G(\kappa t)\,\simeq\,
\displaystyle{
\frac{q_1}{\kappa^2}\,t^{-1}\,+\,q_2\,t \log^3\bigl({\cN}t\bigr)
}\\[0.5cm]
&+&\!\!\!\displaystyle{
\Bigl(n_1\,\log^2\bigl({\cN}t\bigr)\,+\,
n_2\,\log\bigl({\cN}t\bigr)\,+\,n_3\Bigr)\,t
}         \\[0.5cm]
&+&\!\!\!\displaystyle{
\Bigl(n_4\,\log^3\bigl({\cN}t\bigr)\,+\,
n_5\,\log^2\bigl({\cN}t\bigr)\,+\,\bigl(\dots\bigr)\,
\log\bigl({\cN}t\bigr)\,+\,\bigl(\dots\bigr)\Bigr)\,t^3\,+\,\dots\,,
}
\end{array}
\eqno({\rm B}1)
$$
\vskip 0.4cm
$$
\begin{array}{rcl}
k^{-1}\!\!\!&t&\!\!\!{\w Y}_0({\cN}t)\,G(\kappa t)\,\simeq\,
\displaystyle{
\frac{q_1}{\kappa^2}\,t^{-1}\,\log\Bigl(\frac{\ga}2 {\cN} t\Bigr)
}\\[0.5cm]
&+&\!\!\!\displaystyle{
\Bigl(q_2\,\log^4\bigl({\cN}t\bigr)\,+\,
m_1\,\log^3\bigl({\cN}t\bigr)\,+\,m_2\,\log^2\bigl({\cN}t\bigr)\,+\,
m_3\,\log\bigl({\cN}t\bigr)\,+\,m_4 \Bigr)\,t
}\\[0.5cm]
&+&\!\!\!\displaystyle{
\Bigl(m_5\,\log^4\bigl({\cN}t\bigr)\,+\,m_6\,\log^3\bigl({\cN}t\bigr)
\,+\,\dots \Bigr)\,t^3\,+\,\dots\,,
}
\end{array}
\eqno({\rm B}2)
$$
where
$$
\begin{array}{l}
\displaystyle{
n_1\,=\,3\,\log\Bigl(\frac{\kappa}{{\cN}}\Bigr)\,q_2\,+\,q_3\,,
}\\[0.5cm]
\displaystyle{
n_2\,=\,3\,\log^2\Bigl(\frac{\kappa}{{\cN}}\Bigr)\,q_2
\,+\,2\,\log\Bigl(\frac{\kappa}{{\cN}}\Bigr)q_3\,+\,q_4\,,
}   \\[0.5cm]
\displaystyle{
n_3\,=\,q_0\,-\,\frac{{\cN}^2}{\kappa^2}\,\frac{q_1}4\,+\,
\log^3\Bigl(\frac{\kappa}{{\cN}}\Bigr)\,q_2
\,+\,\log^2\Bigl(\frac{\kappa}{{\cN}}\Bigr)q_3
\,+\,\log\Bigl(\frac{\kappa}{{\cN}}\Bigr)q_4\,,
}\\[0.5cm]
\displaystyle{
n_4\,=\,-\,{\cN}^2\,\frac{q_2}4\,+\,\kappa^2\,q_5\,,
}            \\[0.5cm]
\displaystyle{
n_5\,=\,-\,{\cN}^2\,\frac{q_3}4\,+\,\kappa^2\,q_6\,+\,
3\,\log\Bigl(\frac{\kappa}{{\cN}}\Bigr)\,n_4\,,
}            \\[0.5cm]
\end{array}
\eqno({\rm B}3)
$$
and
$$
\begin{array}{l}
\displaystyle{
m_1\,=\,\log\Bigl(\frac{\ga}2\Bigr)\,q_2\,+\,n_1\,,\qquad
m_2\,=\,\log\Bigl(\frac{\ga}2\Bigr)\,n_1\,+\,n_2\,,
}\\[0.5cm]
\displaystyle{
m_3\,=\,\log\Bigl(\frac{\ga}2\Bigr)\,n_2\,+\,n_3\,,\qquad
m_4\,=\,\frac{{\cN}^2}{\kappa^2}\,\frac{q_1}4
\,+\,\log\Bigl(\frac{\ga}2\Bigr)\,n_3\,,
}\\[0.5cm]
\displaystyle{
m_5\,=\,n_4\,,\qquad
m_6\,=\,{\cN}^2\,\frac{q_2}4\,+\,
\log\Bigl(\frac{\ga}2\Bigr)\,n_4\,+\,n_5\,.
}
\end{array}
\eqno({\rm B}4)
$$

Using ({\rm B}1)--({\rm B}4), we pass to the estimation of the 
integrals which enter into $I_{{\cF}}(\rho)$ (4.18). 
First, we obtain:
$$
\begin{array}{rcl}
\displaystyle{
\int\limits_\rho^\infty J_0 ({\cN}t)\,G(\kappa t)\,t\,dt}
&\simeq& \displaystyle{
{\cJ}_0\,+\,{\cJ}_1\,\log\bigl({\cN}\rho\bigr)
}\\[0.5cm]
&+& \displaystyle{
\Bigl({\cJ}_2\,\log^3\bigl({\cN}\rho\bigr)\,+\,
{\cJ}_3\,\log^2\bigl({\cN}\rho\bigr)\,+\,
{\cJ}_4\,\log\bigl({\cN}\rho\bigr)\,+\,
{\cJ}_5\Bigr)\,\rho^2
}      \\[0.5cm]
&+& \displaystyle{
\Bigl({\cJ}_6\,\log^3\bigl({\cN}\rho\bigr)\,+\,
{\cJ}_7\,\log^2\bigl({\cN}\rho\bigr)\,+\,\dots \Bigr)\,\rho^4\,,
}
\end{array}
\eqno({\rm B}5)
$$
where
$$
\begin{array}{l}
\displaystyle{
{\cJ}_0\,=\,{\cI}_J\,,\qquad {\cJ}_1\,=\,-k\,\frac{q_1}{\kappa^2}\,,
\qquad {\cJ}_2\,=\,-k\,\frac{q_2}2\,,
}         \\[0.5cm]
\displaystyle{
{\cJ}_3\,=\,k\,\Bigl(\frac{3 q_2}4\,-\,\frac{n_1}2 \Bigr)\,,
\qquad {\cJ}_4\,=\,k\,\Bigl(-\frac{3 q_2}4\,+\,
\frac{n_1\,-\,n_2}2 \Bigr)\,,
}\\[0.5cm]
\displaystyle{
{\cJ}_5\,=\,k\,\Bigl(\frac{3 q_2}8\,+\,
\frac{n_2\,-\,n_1}4\,-\,\frac{n_3}2 \Bigr)\,,
\qquad {\cJ}_6\,=\,-k\,\frac{n_4}4\,,}\\[0.5cm]
\displaystyle{
{\cJ}_7\,=\,k\,\Bigl(\frac{3 n_4}{16}\,-\,
\frac{n_5}4 \Bigr)\,.
}
\end{array}
\eqno({\rm B}6)
$$
Besides, dots in (B5) imply terms proportional to the first and
zeroth powers of $\log\bigl({\cN}\rho\bigr)$. The constant 
${\cI}_J$ which gives ${\cJ}_0$ will be presented below.
Further, we obtain:
$$
\begin{array}{rcl}
\displaystyle{
\int\limits_\rho^\infty {\w Y}_0 ({\cN}t)\,G(\kappa t)\,t\,dt}
\!\!\!&\simeq&\!\!\! \displaystyle{
{\cY}_0\,+\,{\cY}_1\,\log^2\bigl({\cN}\rho\bigr)
\,+\,{\cY}_2\,\log\bigl({\cN}\rho\bigr)
}\\[0.5cm]
+\,\Bigl({\cY}_3\,\log^4\bigl({\cN}\rho\bigr)\!\!\!&+&\!\!\!
{\cY}_4\,\log^3\bigl({\cN}\rho\bigr)\,+\,
{\cY}_5\,\log^2\bigl({\cN}\rho\bigr)\,+\,
{\cY}_6\,\log\bigl({\cN}\rho\bigr)\,+\,
{\cY}_7\Bigr)\,\rho^2
 \\[0.5cm]
+\, 
\Bigl({\cY}_8\,\log^4\bigl({\cN}\rho\bigr)\!\!\!&+&\!\!\!
{\cY}_9\,\log^3\bigl({\cN}\rho\bigr)\,+\,\dots \Bigr)\,\rho^4\,,
\end{array}
\eqno({\rm B}7)
$$
where
$$
\begin{array}{l}
\displaystyle{
{\cY}_0\,=\,{\cI}_Y\,,\quad 
           {\cY}_1\,=\,-k\,\frac{q_1}{2 \kappa^2}\,,\quad
{\cY}_2\,=\,-k\,\frac{q_1}{\kappa^2}\,\log\frac{\ga}2\,,\quad
{\cY}_3\,=\,-k\,\frac{q_2}{2}\,,
}                              \\[0.5cm]
\displaystyle{
{\cY}_4\,=\,k\,\Bigl(q_2\,-\,\frac{m_1}2\Bigr)\,,\quad
{\cY}_5\,=\,k\,\Bigl(-\frac{3 q_2}2\,+\,\frac{3 m_1}4\,-\,
\frac{m_2}2\Bigr)\,,
}\\[0.5cm]
\displaystyle{
{\cY}_6\,=\,k\,\Bigl(\frac{3 q_2}2\,-\,\frac{3 m_1}4\,+\,
\frac{m_2\,-\,m_3}2\Bigr)\,,
}\\[0.5cm]
\displaystyle{
{\cY}_7\,=\,k\,\Bigl(-\frac{3 q_2}4\,+\,\frac{3 m_1}8\,+\,
\frac{m_3\,-\,m_2}4\,-\,\frac{m_4}2\Bigr)\,,
}\\[0.5cm]
\displaystyle{
{\cY}_8\,=\,-k\,\frac{m_5}{4}\,,\quad
{\cY}_9\,=\,k\,\frac{m_5\,-\,m_6}4\,,
}
\end{array}
\eqno({\rm B}8)
$$
and dots in (B7) corresponds to the second, first, and
zeroth powers of $\log\bigl({\cN}\rho\bigr)$.

With the expansions ({\rm B}5) and ({\rm B}7) at hands, 
we pass to the products we are interested in to express 
$I_{{\cF}}$ (4.18):
$$
\begin{array}{rcl}
\displaystyle{
{\w Y}_0 ({\cN}\rho)
\int\limits_\rho^\infty J_0 ({\cN}t)\,G(\kappa t)\,t\,dt}
\!\!\!&=&\!\!\! \displaystyle{
{\h Y}_0\,+\,{\h Y}_1\,\log^2\bigl({\cN}\rho\bigr)
\,+\,{\h Y}_2\,\log\bigl({\cN}\rho\bigr)
}\\[0.5cm]
+\,\Bigl({\h Y}_3\,\log^4\bigl({\cN}\rho\bigr)\!\!\!&+&\!\!\!
{\h Y}_4\,\log^3\bigl({\cN}\rho\bigr)\,+\,
{\h Y}_5\,\log^2\bigl({\cN}\rho\bigr)\,+\,
{\h Y}_6\,\log\bigl({\cN}\rho\bigr)\,+\,
{\h Y}_7\Bigr)\,\rho^2
 \\[0.5cm]
+\,\Bigl({\h Y}_8\,\log^4\bigl({\cN}\rho\bigr)\!\!\!&+&\!\!\!
{\h Y}_9\,\log^3\bigl({\cN}\rho\bigr)\,+\dots\Bigr)\,\rho^4\,,
\end{array}
\eqno({\rm B}9)
$$
where
$$
\begin{array}{l}
\displaystyle{
{\h Y}_0\,\equiv\,\log\Bigl(\frac{\ga}2\Bigr)\,{\cI}_J\,,\quad
{\h Y}_1\,=\,{\cJ}_1\,,\quad{\h Y}_2\,=\,{\cJ}_0\,+\,
\log\Bigl(\frac{\ga}2\Bigr)\,{\cJ}_1\,,\quad
{\h Y}_3\,=\,{\cJ}_2\,, 
}\\[0.5cm]
\displaystyle{
{\h Y}_4\,=\,{\cJ}_3\,+\,\log\Bigl(\frac{\ga}2\Bigr)\,{\cJ}_2\,,\quad
{\h Y}_5\,=\,-\frac{{\cN}^2}4\,{\cJ}_1\,+\,
\log\Bigl(\frac{\ga}2\Bigr)\,{\cJ}_3\,+\,{\cJ}_4\,,
}\\[0.5cm]
\displaystyle{
{\h Y}_6\,=\,\frac{{\cN}^2}4\,\left(-\,{\cJ}_0\,+\,
\Bigl(1\,-\,\log\frac{\ga}2\Bigr){\cJ}_1\right)\,+\,
\log\Bigl(\frac{\ga}2\Bigr)\,{\cJ}_4\,+\,{\cJ}_5\,,
}\\[0.5cm]
\displaystyle{
{\h Y}_7\,=\,\frac{{\cN}^2}4\,\Bigl(1\,-\,\log\frac{\ga}2\Bigr){\cJ}_0
\,+\,\log\Bigl(\frac{\ga}2\Bigr)\,{\cJ}_5\,,\quad
{\h Y}_8\,=\,-\,\frac{{\cN}^2}4\,{\cJ}_2\,+\,{\cJ}_6\,,
}\\[0.5cm]
\displaystyle{
{\h Y}_9\,=\,\frac{{\cN}^2}4\,\Bigl(1\,-\,
           \log\frac{\ga}2\Bigr){\cJ}_2
\,-\,\frac{{\cN}^2}4\,{\cJ}_3\,+\,\log\Bigl(\frac{\ga}2\Bigr)\,
{\cJ}_6\,+\,{\cJ}_7\,.}
\end{array}
\eqno({\rm B}10)
$$
Now we obtain the following expansions:
$$
\begin{array}{rcl}
\displaystyle{
J_0 ({\cN}\rho)
\int\limits_\rho^\infty {\w Y}_0 ({\cN}t)\,G(\kappa t)\,t\,dt}
\!\!\!&=&\!\!\! \displaystyle{
{\h J}_0\,+\,{\h J}_1\,\log^2\bigl({\cN}\rho\bigr)
\,+\,{\h J}_2\,\log\bigl({\cN}\rho\bigr)
}\\[0.5cm]
+\,\Bigl({\h J}_3\,\log^4\bigl({\cN}\rho\bigr)\!\!\!&+&\!\!\!
{\h J}_4\,\log^3\bigl({\cN}\rho\bigr)\,+\,
{\h J}_5\,\log^2\bigl({\cN}\rho\bigr)\,+\,
{\h J}_6\,\log\bigl({\cN}\rho\bigr)\,+\,
{\h J}_7\Bigr)\,\rho^2
 \\[0.5cm]
+\,\Bigl({\h J}_8\,\log^4\bigl({\cN}\rho\bigr)\!\!\!&+&\!\!\!
{\h J}_9\,\log^3\bigl({\cN}\rho\bigr)\,+\dots \Bigr)\,\rho^4\,,
\end{array}
\eqno({\rm B}11)
$$
where
$$
\begin{array}{l}
\displaystyle{
{\h J}_0\,\equiv\,{\cI}_Y\,,\quad
{\h J}_i\,=\,{\cY}_i\,,\quad i\,=\,1\,, 2\,, 3\,, 4\,,
}\\[0.5cm]
\displaystyle{
{\h J}_5\,=\,-\frac{{\cN}^2}4\,{\cY}_1\,+\,{\cY}_5\,,\qquad
{\h J}_6\,=\,-\frac{{\cN}^2}4\,{\cY}_2\,+\,{\cY}_6\,,
}\\[0.5cm]
\displaystyle{
{\h J}_7\,=\,-\frac{{\cN}^2}4\,{\cY}_0\,+\,{\cY}_7\,,
}\\[0.5cm]
\displaystyle{
{\h J}_8\,=\,-\frac{{\cN}^2}4\,{\cY}_3\,+\,{\cY}_8\,,\qquad
{\h J}_9\,=\,-\frac{{\cN}^2}4\,{\cY}_4\,+\,{\cY}_9\,.
}
\end{array}
\eqno({\rm B}12)
$$

Now we sum up expansions ({\rm B}9) and ({\rm B}11):
$$
\begin{array}{l}
\displaystyle{
I_{{\cF}}(\rho)\,=\,-\,{\w Y}_0 ({\cN}\rho)
\int\limits_\rho^\infty J_0 ({\cN}t)\,G(\kappa t)\,t\,dt\,+\,
J_0 ({\cN}\rho)
\int\limits_\rho^\infty {\w Y}_0 ({\cN}t)\,G(\kappa t)\,t\,dt}
\,=\\[0.5cm]
=\,
\displaystyle{
{\h J}_0\,-\,{\h Y}_0\,+\,
\bigl({\h J}_1\,-\,{\h Y}_1\bigr)\,\log^2\bigl({\cN}\rho\bigr)
\,+\,\bigl({\h J}_2\,-\,{\h Y}_2\bigr)\,\log\bigl({\cN}\rho\bigr)
}\\[0.5cm]
+\,\Bigl(
\bigl({\h J}_4\,-\,{\h Y}_4\bigr)\,
         \log^3\bigl({\cN}\rho\bigr)\,+\,
\bigl({\h J}_5\,-\,{\h Y}_5\bigr)\,
         \log^2\bigl({\cN}\rho\bigr)\,+\,
\bigl({\h J}_6\,-\,{\h Y}_6\bigr)\,\log\bigl({\cN}\rho\bigr)\,+\,
{\h J}_7\,-\,{\h Y}_7\Bigr)\,\rho^2
 \\[0.5cm]
+\,\Bigl(\bigl({\h J}_9\,-\,{\h Y}_9\bigr)\,
           \log^3\bigl({\cN}\rho\bigr)\,+\,\dots
\Bigr)\,\rho^4\,,
\end{array}
\eqno({\rm B}13)
$$
where
$$
\begin{array}{l}
\displaystyle{
{\h J}_0\,-\,{\h Y}_0\,=\,{\cI}_Y\,-\,
\log\Bigl(\frac{\ga}2\Bigr){\cI}_J\,,
}\\[0.5cm]
\displaystyle{
{\h J}_1\,-\,{\h Y}_1\,=\,k\,\frac{q_1}{2\kappa^2}
\,=\,k\,\frac{p_1}{8\kappa^2}\,,
}\\[0.5cm]
\displaystyle{
{\h J}_2\,-\,{\h Y}_2\,=\,-{\cI}_J\,,\qquad
{\h J}_4\,-\,{\h Y}_4\,=\,k\,\frac{q_2}4\,=\,k\,\frac{p_2}{24}\,,
}\\[0.5cm]
\displaystyle{
{\h J}_5\,-\,{\h Y}_5\,=\,k
\biggl(-\frac{{\cN}^2}{\kappa^2}\,\frac{q_1}8\,-\,
\Bigl(1\,-\,\log\frac{\kappa}{{\cN}}\Bigr)
\,\frac{3 q_2}4\,+\,\frac{q_3}4\biggr)\,=
}\\[0.5cm]
\hskip 2.0cm \displaystyle{
=\,k
\biggl(\Bigl(1\,-\,\frac{{\cN}^2}{\kappa^2}\Bigr)\,
\frac{p_1}{32}\,-\,
\Bigl(1\,-\,\log\frac{\kappa}{{\cN}}\Bigr)
\,\frac{p_2}8\,+\,\frac{p_3}8\biggr)\,,
}\\[0.5cm]
\displaystyle{
{\h J}_6\,-\,{\h Y}_6\,=\,\frac{{\cN}^2}4\,
\left({\cJ}_0\,-\,\Bigl(1\,-\,\log\frac{\ga}2\Bigr){\cJ}_1\,-\,
{\cY}_2\right)\,-\,\log\Bigl(\frac{\ga}2\Bigr)\,{\cJ}_4\,-
}\\[0.5cm]
\hskip 2.0cm
-\,{\cJ}_5\,+\,{\cY}_6\,,\\[0.5cm]
\displaystyle{
{\h J}_7\,-\,{\h Y}_7\,=\,-\,\frac{{\cN}^2}4\,
\left({\cY}_0\,+\,\Bigl(1\,-\,\log\frac{\ga}2\Bigr){\cJ}_0 \right)
\,-\,\log\Bigl(\frac{\ga}2\Bigr)\,{\cJ}_5\,+\,{\cY}_7\,,
}\\[0.5cm]
\displaystyle{
{\h J}_9\,-\,{\h Y}_9\,=\,k\,\biggl(-
\frac{{\cN}^2}{64}\,q_2\,+\,\frac{\kappa^2}{16}\,q_5\biggr)
\,=\,k\,\frac{\kappa^2\,-\,{\cN}^2}{384}\,p_2\,.
}
\end{array}
\eqno({\rm B}14)
$$
The terms corresponding to ${\h J}_3-{\h Y}_3$ and 
${\h J}_8-{\h Y}_8$ do not appear since ${\cY}_3={\cJ}_2$
and $m_5=n_4$.

Eventually, it is necessary to re-arrange the series ({\rm B}13) as 
follows:
$$
\begin{array}{l}
I_{{\cF}}(\rho)\,=\,
\displaystyle{
{\w {\cI}_Y}\,-\,\log\Bigl(\frac{\ga}2{\cN}\Bigr){\w {\cI}_J}
\,+\,k\,\frac{q_1}{2\kappa^2}\,\log^2\rho
\,-\,{\w {\cI}_J}\,\log\rho
\,+\,k\,\frac{q_2}4\,\rho^2\log^3\rho\,+
}\\[0.5cm]
+\,\Bigl(
\bigl({\h J}_4\,-\,{\h Y}_4\bigr)\,
         \log\bigl({\cN}^3\bigr)\,+\,
{\h J}_5\,-\,{\h Y}_5\Bigr)\,\rho^2\log^2\rho
\\[0.5cm]
+\,\Bigl(
\bigl({\h J}_4\,-\,{\h Y}_4\bigr)\,
        3\,\log^2{\cN}\,+\,
\bigl({\h J}_5\,-\,{\h Y}_5\bigr)\,
         \log\bigl({\cN}^2\bigr)\,+\,
{\h J}_6\,-\,{\h Y}_6\Bigr)\,\rho^2\log\rho
\\[0.5cm]
+\,\Bigl(
\bigl({\h J}_4\,-\,{\h Y}_4\bigr)\,
         \log^3{\cN}\,+\,
\bigl({\h J}_5\,-\,{\h Y}_5\bigr)\,
         \log^2{\cN}\,+\,
\bigl({\h J}_6\,-\,{\h Y}_6\bigr)\,\log {\cN}\,+\,
{\h J}_7\,-\,{\h Y}_7\Bigr)\,\rho^2
\\[0.5cm]
+\,\Bigl({\h J}_9\,-\,{\h Y}_9\Bigr)\,\rho^4\log^3\rho\,,
\end{array}
\eqno({\rm B}15)
$$
where ${\h J}_i-{\h Y}_i$, $i=4,\dots , 9$ are
given by (B14) (where (B6) and (B8) must be used) and the following 
notations are adopted:
$$
\begin{array}{l}
\displaystyle{
{\cI}_J\,\equiv\,{\w {\cI}_J}\,+\,k\,
     \frac{q_1}{\kappa^2}\log{\cN}}\,,\\[0.5cm]
\displaystyle{
{\cI}_Y\,\equiv\,{\w {\cI}_Y}\,+\,k\,\frac{q_1}{\kappa^2}\log{\cN}\,
\biggl(\frac{\log {\cN}}2\,+\,\log\frac{\ga}2\biggr)
     }\,,
\end{array}
\eqno({\rm B}16)
$$
and
$$
\begin{array}{l}
\displaystyle{
{\w {\cI}_J}\,\equiv\,\lim_{\ep\to 0+}
\left(
\int\limits_\ep^\infty J_0 ({\cN}t)\,G(\kappa t)\,t\,dt
\,-\,k\,\frac{q_1}{\kappa^2}\int\limits_\ep^1
\frac{dt}{t}\right)\,,}\\[0.5cm]
\displaystyle{
{\w {\cI}_Y}\,\equiv\,\lim_{\ep\to 0+}
\left(
\int\limits_\ep^\infty {\w Y}_0 ({\cN}t)\,G(\kappa t)\,t\,dt
\,-\,k\,\frac{q_1}{\kappa^2}\int\limits_\ep^1
\log\Bigl(\frac{\ga}2\,{\cN}t\Bigr)\frac{dt}{t}
\right)\,.}
\end{array}
\eqno({\rm B}17)
$$

\end{document}